\documentclass[preprint,12pt]{elsarticle}

\usepackage{tikz}
\usepackage{amssymb}
\usepackage{amsmath}
\usepackage{booktabs}
\usepackage{graphicx}
\usepackage{caption}
\usepackage{subcaption}
\usepackage{url}
\usepackage{tcolorbox}
\usepackage[superscript,biblabel]{cite}

\newtheorem{theorem}{Theorem}{\bf}{}
\newtheorem{lemma}{Lemma}{\bf}{}
{\bf}{}
{\bf}{}
{\bf}{}
{\bf}{}
{\bf}{}

\journal{-}

\begin{document}

\begin{frontmatter}



\title{The Role of Asymptomatic Individuals in the COVID-19 Pandemic via Complex Networks}


\author{Leonardo~Stella, Alejandro~Pinel~Mart\'inez, Dario~Bauso and Patrizio~Colaneri}

\address{L. Stella is with the Department of Computing, College of Science \& Engineering, University of Derby, Kedleston Road, DE22 1GB, United Kingdom, e-mail: (l.stella@derby.ac.uk).}

\address{A. Pinel Mart\'inez is with the he Department of Computing, College of Science \& Engineering, University of Derby, Kedleston Road, DE22 1GB, United Kingdom, e-mail: (a.pinelmartinez1@unimail.derby.ac.uk).}

\address{D. Bauso is with the Jan C. Willems Center for Systems and Control, ENTEG, Faculty of Science and Engineering, University of Groningen, The Netherlands, and with the Dipartimento di Ingegneria, University of Palermo, Italy, e-mail: (d.bauso@rug.nl).}

\address{P. Colaneri is with Dipartimento di Elettronica e Informazione, Politecnico di Milano, Italy, e-mail: (patrizio.colaneri@polimi.it).}

\end{frontmatter}



{\section*{\LARGE Summary}
\section*{Background}
Recent seroprevalence studies have tried to estimate the real number of asymptomatic cases affected by COVID-19. It is of paramount importance to understand the impact of these infections in order to prevent a second wave. This study aims to model the interactions in the population by means of a complex network and to shed some light on the effectiveness of localised control measures in Italy in relation to the school opening in mid-September.}

{\section*{Methods}
The formulation of an epidemiological predictive model is given: the advantage of using this model lies in that it discriminates between asymptomatic and symptomatic cases of COVID-19 as the interactions with these two categories of infected individuals are captured separately, allowing for a study on the impact of asymptomatic cases. This model is then extended to a structured nonhomogeneous version by means of the Watts-Strogatz complex network, which is adopted widely to model societal interactions as it holds the small world property. Finally, a case study on the situation in Italy is given: first the homogeneous model is used to compare the official data with the data of the recent seroprevalence study from Istat; second, in view of the return to school in mid-September, a study at regional level is conducted.}

{\section*{Findings}
The proposed model captures an aspect of COVID-19 which is crucial in controlling the further spread of the contagion and in preventing a second wave: {the interactions with undetected cases who have therefore not been isolated}, mostly asymptomatic or paucisymptomatic, namely 1-2 symptoms without anosmia or ageusia, and their impact on latent infections. The evolution of the pandemic is well captured by model fitting with official data from the Protezione Civile, both at a national level as well as regional level. The case study provides insight on the potential effects of localised restrictions, without the coordination at a national level. The results of this study highlight the importance of coordinating the deployment of appropriate control measures that take into account the role of asymptomatic infections, especially in younger individuals, and inter-regional connectivity in Italy.}

{\section*{Interpretation}
It is vital to adhere to the prescribed public health measures in order to delay the spread of SARS-CoV-2 and prevent a second wave with potential disruptive national lockdowns. Asymptomatic infected are estimated to be around a third of the official data, but their real number can be much higher if the individuals with one or two minor symptoms are included. The results emphasise the need for coordinated control measures that account for the interactions among different regions in Italy, and potentially among the countries in Europe.}

{\section*{Funding}
None.}

{\section*{\LARGE Introduction}}
The COVID-19 respiratory syndrome, associated with the novel strand of Coronavirus called SARS-CoV-2, has had a massive impact worldwide. Initially found in Wuhan, in the heart of Hubei Province, China\cite{Zhu_2020}, it has quickly spread since last December to almost every country in the world, with the most affected being the US, Spain, UK, Italy, France, Germany, Russia, Turkey, Iran, and China. This has caused severe consequences and a large number of deaths, mostly due to the ease of transmission, i.e. the virality, of this disease.
For an infectious disease outbreak such as the one caused by COVID-19, predictive mathematical models play an important role for the planning of effective control strategies. Among the models formulated over the years\cite{Anderson_1992,Hethcote_2000}, the susceptible-infected-recovered model (SIR) is possibly one of the most used epidemiological models: the population is split into three stages of infection, sometimes called compartments, thus the terminology \emph{compartmental models}, as reported in an early work by Kermack and McKendrick in 1927\cite{Kermack_1927}. A variant of these classic compartmental models used to tackle the specific features of SARS can be found in the work of Gumel et al.\cite{Gumel_2004} and similar equations can be found in the framework developed for the HIV transmission in heterogeneous populations\cite{May_1988}.

Several aspects of this virus have been investigated: some research assessed the effectiveness of different response strategies\cite{Zhang_2020}, other authors focused on modelling the various stages of the disease and the death rate, as done in a seminal work by Giordano et al.\cite{Giordano_2020} Early research in China showed unique epidemiological traits of the COVID-19 virus\cite{Wang_2020}, most notably the fact that a large portion of transmissions were caused by asymptomatic individuals, whether they were showing mild or no symptoms at all. Indeed, further research demonstrated that asymptomatic and symptomatic individuals have the same viral load and thus the same capability to further spread the virus\cite{Zou_2020}, and the work of Rothe et al. for a case of transmission from an asymptomatic individual in Germany\cite{Rothe_2020}. In the context of data driven models, Bertozzi et al. find a relation between branching point processes and classical compartmental models such as susceptible-infected-recovered (SIR) and susceptible-exposed-infected-recovered (SEIR), whilst fitting the models with data from a variety of countries, including China, Italy, Japan, and other countries\cite{Bertozzi_2020}. A study that investigates whether daily test reports can help authorities to control the epidemic\cite{Casella_2020}, discusses how mitigation strategies can fail when modelled because of various factors, such as delay, unstable dynamics, and uncertainty in the feedback loop. For the Italian situation, the work of Della Rossa et al. provides interesting insight on the need to coordinate the efforts in controlling the situations in an inter-regional setting, and highlights the need of such coordination by means of a network model\cite{DellaRossa_2020}. In the work of Yilmaz et al., the authors discuss how to identify and analyze bridges between communities in graphs with the purpose to understand how to track and where to start tests on which individuals\cite{Yilmaz_2020}. {Another study includes a particle-based mean field model that investigates the pros and cons of social distancing through an approach that compares individuals to molecules in a solution\cite{Franco_2020}. {A very early model of this disease was given in the work by Calafiore et al., where the novelty lies in including a proportionality factor in a standard SIR model to account for hidden infections\cite{Calafiore_2020}.} In a model on the case for the UK, the authors account for four main elements and a finer level of detail for each of them in assessing the impact of the speed in which the immunity is lost\cite{Friston_2020}. A risk sensitivity analysis on the economic impact of the disease where the optimising behaviour of agents to influence future transitions is considered by the work by Garibaldi et al.\cite{Garibaldi_2020}.} The work of Pastor-Satorras provides a survey of the literature on complex networks for epidemic processes\cite{Satorras_2015}{, and applications of complex networks to epidemic processes in evolutionary dynamics can be found in the work by Tan et al.\cite{Tan_2014}}

\begin{tcolorbox}[width=\textwidth,colback={white},title={-},outer arc=0mm,colupper=black]    
\textbf{Research in context}\\
\textbf{Evidence before this study}\\
Italy has been the first European country to face the challenges brought by the COVID-19 pandemic. Predictive mathematical models are a useful tool to understand the proportion of this pandemic and consider different control measures, but few have considered the impact of asymptomatic in a structured setting, where the structure is captured by complex networks. Our search spanned from medical journals from PubMed in general and specific sources in control and data driven modelling until August 31, 2020, using the terms ``control''  and ``data driven modelling'', without date or language restrictions. As for the databases, we used the official data from the Italian Ministry of Health and the Istat seroprevalence study. Seroprevalence studies are recent in many European countries and many works considered them in modelling the evolution of SARS-CoV-2. The impact of asymptomatic infected must be understood to prevent a second wave of this virus.

\textbf{Added value of this study}\\
This is one of the first studies that uses data from the Istat seroprevalence analysis to model the evolution of SARS-CoV-2 at national and regional levels in Italy with emphasis on modelling the interactions between susceptible individuals and asymptomatic infected. Our study proposes a predictive model, inspired by classical epidemiological models, that emphasises the impact of asymptomatic infected in spreading the disease through a measure of the interactions in the population. This study analyses the possible scenarios following the return to schools in mid-September by capturing the heterogeneity in the population by means of complex networks, graphs that mimic real features present in biological, computer and social networks. Finally, our case study for the Italian case confirms that only a coordinated policy at national and regional levels can be effective in controlling the virus in the forthcoming months.

\textbf{Implications of all the available evidence}\\
Our study can serve as a reference to study the situation in other contexts and countries, suggesting similar outcomes for a multi-national approach in the European case or in the US. Because of the present difficulty to reach herd immunity without counting a large number of deaths due to the inability of health systems to face the speed in which this disease spreads, our results suggest, in line with previous evidence and  data driven models, that the only way to monitor this disease is through coordinated action and monitoring of the asymptomatic or paucisymptomatic infected. 

\end{tcolorbox}

{\section*{\LARGE Methods}}\label{sec:model}
{\section*{\large Epidemic Model}}
In this section, we present the formulation of the model {that} we propose, which takes inspiration from notorious compartmental models such as the widely used susceptible-infected-recovered (SIR) model{, and more precisely from the susceptible-exposed-infected-recovered (SEIR) model. The peculiarity of a compartmental model is that} the population is divided into a discrete set of states and individuals can only exist in these states: for instance, in the SIR model, individuals can be susceptible to the virus (healthy), contract the virus and thus be infected, and finally be immune to the virus when they recover. This model accounts for those diseases that do provide long term immunity to future infections from the same virus {through the presence of antibodies in the host organism}.

We named our model \emph{SAIR}, because of the state variables we chose to include: Susceptible, Asymptomatic infected, symptomatic Infected and Removed. We choose to use the term \emph{removed} in place of the more common \emph{recovered} because we do not discriminate between individuals who recover from the disease and individuals who passed away. The term \emph{removed} is also used commonly in the literature, and in important contributions in the field\cite{Moreno_2002}. As previously mentioned, our model is a variant of the susceptible-exposed-infected-recovered (SEIR)\cite{Pastor-Satorras_2015}, but with notable differences:
\begin{itemize}
\item Our study focuses on the impact of the undetected asymptomatic individuals in spreading the virus. Some of these can show symptoms at a later stage of the disease, and we assume that in an initial stage no individuals show symptoms.
\item We provide an estimation of the parameters of infection through a case study for Italy. We estimate the ratio between asymptomatic and symptomatic infected and support our work with the estimate from the Istat seroprevalence study\cite{Istat_2020}.
\item We investigate the impact of the lockdown measures in controlling the spread of the virus, by modelling the frequency of contacts among the individuals in the population {via an average number of contacts, first, and then through the small-world} complex network {model}.
\end{itemize}

The susceptible-asymptomatic-infected-removed (SAIR) model that we present in the following is a discrete-state continuous-time system. In a first approximation, individuals are considered homogeneous, namely they share the same properties when in the same state (or compartment). The state variables of the model represent the densities of susceptible, asymptomatic infected, symptomatic infected and removed individuals. These quantities are denoted by $S(t)$,  $A(t)$, $I(t)$ and $R(t)$, respectively. Each state variable belongs to $\mathbb{R}_0^+$. In the mean-field limit, the following system of ODEs describes the time evolution of the population:
\begin{equation} \label{eq:modSAIR}
\left\{\begin{array}{ll}
\dot S(t) = -S(t)(\bar k_1 \gamma A(t) + \bar k_2 \lambda I(t)), \\
\dot A(t) = S(t)(\bar k_1 \gamma A(t) + \bar k_2 \lambda I(t)) - A(t)(\alpha + \sigma), \\
\dot I(t) = \alpha A(t) - \mu I(t), \\
\dot R(t) = \sigma A(t) + \mu I(t), \\
\end{array} \right.
\end{equation}
where the uppercase Latin letters represent the known densities, $\bar k_1$ and $\bar k_2$ describe the average number of contacts per unit time with asymptomatic and symptomatic infected individuals, respectively, and the lowercase Greek letters represent the parameters of the system. In particular, these parameters are positive quantities and have the following physical interpretation: $\gamma$ and $\lambda$ denote the microscopic transmission rate, the former due to contacts between a susceptible person and an asymptomatic infected, the latter due to contacts between a susceptible person and a symptomatic infected; infected individuals decay into the removed class at rate $\sigma$ from the asymptomatic infected state and at rate $\mu$ from the symptomatic infected state, respectively; finally, $\alpha$ is the rate in which asymptomatic individuals start to have symptoms.

System (\ref{eq:modSAIR}) is a nonlinear positive system, more precisely it is bilinear, since the highest degree that we have is at most {two, namely the multiplication} between two state variables. The fact that the system is positive means that, given an initial condition $S(0), A(0), I(0), R(0)  \ge 0$, all the state variables take nonnegative values for $t \ge 0$. Furthermore, through the conservation of mass, namely $\dot S(t) + \dot A(t) + \dot I(t) + \dot R(t) = 0$, all state variables are linked through the normalisation condition:
$$S(t) + A(t) + I(t) + R(t) = 1,$$
meaning that the sum of  {all} the state variables is constant at any given time {and equal to one}. 

\begin{figure}[t]
\begin{centering}
      \begin{tikzpicture} 
      \tikzstyle{mystyle}=[circle,draw,thick,minimum size=2cm,draw=black,fill=white,align=center]
      \path 	(-3.75,0) node[mystyle](a) {{\Large \color{red} S}\\{\scriptsize {\normalsize \color{red} S}usceptible}}
      		(0,1.75) node[mystyle](b) {{\Large \color{red} A}\\{\scriptsize {\normalsize \color{red} A}symptomatic}\\{\scriptsize Infected}}
		(0,-1.75) node[mystyle](c) {{\Large \color{red} I}\\{\scriptsize Symptomatic}\\{\scriptsize {\normalsize \color{red} I}nfected}}
		(3.75,0) node[mystyle](d) {{\Large \color{red} R}\\{\scriptsize {\normalsize \color{red} R}emoved}};
	\draw[->,orange,very thick] (a) .. controls +(1.25cm,1.25cm) and +(left:2cm) .. node[left] {\large  $\gamma A + \lambda I$} (b);
	\draw[->,red,very thick] (b) .. controls +(down:2cm) and +(up:2cm) .. node[left] {\large  $\alpha$} (c);
	\draw[->,teal,very thick] (b) .. controls +(right:2cm) and +(-.75cm,.75cm) .. node[above] {\large  $\sigma$} (d);
	\draw[->,teal,very thick] (c) .. controls +(right:2cm) and +(-.75cm,-.75cm) .. node[below] {\large  $\mu$} (d);
      \end{tikzpicture}
\caption{Markov chain representation describing the transition rates between the states of the SAIR model in~(\ref{eq:modSAIR}).}
\label{fig:markov}
\end{centering}
\end{figure}
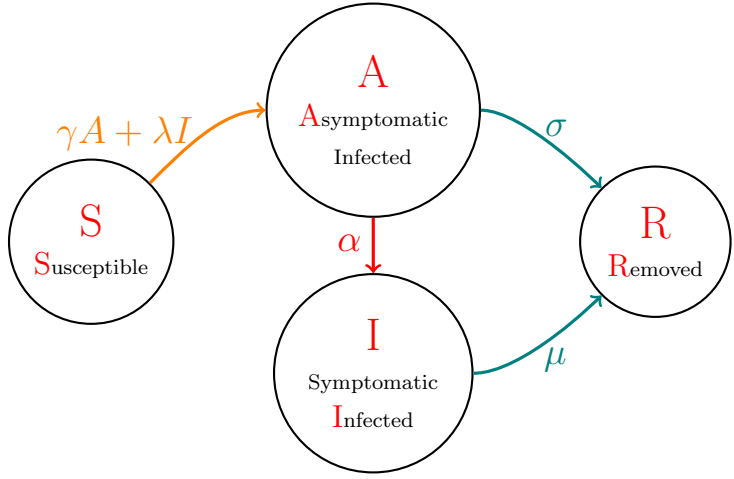 

A few considerations follow. In the work of Giordano et al.\cite{Giordano_2020}, $\gamma > \lambda$, due to the fact that people are more likely in contact with, or closer to, asymptomatic infected individuals rather than with people that show clear symptoms. In our case, we model the interactions among individuals via $\bar k_1$ and $\bar k_2$, and it is obvious to conclude that, in general, $\bar k_1 > \bar k_2$ for the same reason. Finally, in our model we assume the \emph{homogeneous mixing} hypothesis\cite{Anderson_1992}, which asserts that the rate of infection per capita of the susceptible individuals is proportional to the number of people already infected. Because of this hypothesis, system (\ref{eq:modSAIR}) is treated as a mean-field model where the rate of contacts between susceptibles and both symptomatic and asymptomatic {individuals} is assumed constant, independently of any source of heterogeneity present in the system. Figure \ref{fig:markov} depicts the Markov chain corresponding to system (\ref{eq:modSAIR}).

\begin{figure}[t]
\centering
	\begin{tikzpicture}
	\begin{scope}[shift={(-1,0)},scale=1.5]
		\draw [thick] (0.2,-0.25) rectangle (2,0.75);
		\node at (1.1,0.25) {\Huge $\Sigma$};
		
		\draw [thick] (0.2,-2.0) rectangle (0.2+1.8,-2.0+1.0);
		\node at (.2+.9,-2+.5) {\Large $S(t)$};
		
		\draw [->,  thick]  (0.2,-1.5) -- (-.75,-1.5) -- (-.75,.3) -- (.2,.3);
		\draw [->,  thick] (0.5,1.4) -- (0.5,0.75);
		\node at (.9,1.2) {\large $x(0)$};
		
		\draw [->, thick] (2.8,.3) -- (2.8,-1.5) -- (2.0,-1.5);
		\draw [thick,fill=black] (2.8,.3) circle (1pt);
		
		
		\draw [->, thick] (2,.3) -- (3.4,.3);
		\node at (2.5,0.5) {\large $y(t)$};
	\end{scope}
	\end{tikzpicture}
	\caption{The SAIR system in feedback form corresponding to equations in~(\ref{eq:feedback1})-(\ref{eq:feedback3}), where the subsystem indicated by $\Sigma$ can be seen as a positive linear system under feedback, and we can calculate the values of $R$ based on its output.}
\label{fig:feedback}
\end{figure}
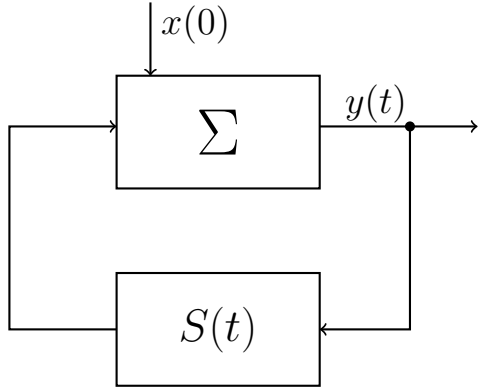

Let $z(t) = [S(t)\, A(t)\, I(t)\, R(t)]^T$, system (\ref{eq:modSAIR}) can be rewritten in matrix form as:
$$\dot z(t) = G z(t),$$
which is equivalent to
\begin{equation} \label{eq:mfSAIR}
\underbrace{\left[ \begin{array}{c}
\dot S \\
\dot A \\
\dot I \\
\dot R \end{array} \right]}_{\dot z} = 
\underbrace{\left[ \begin{array}{cccc}
0 & -\bar k_1\gamma S & -\bar k_2\lambda S & 0 \\
0 & \bar k_1\gamma S - \alpha - \sigma & \bar k_2\lambda S & 0 \\
0 & \alpha & - \mu & 0 \\
0 & \sigma & \mu & 0 \\ 
\end{array} \right]}_{G} 
\underbrace{\left[ \begin{array}{c}
S \\
A \\
I \\
R \end{array} \right]}_{z},
\end{equation}
where the dependence on time is implicit, e.g. $S := S(t)$, for the sake of brevity. As depicted in Fig.~\ref{fig:feedback}, the above system can be rewritten in feedback form, where the subsystem consisting of variables $A$ and $I$ can be seen as a positive linear system under feedback. Let $x(t) = [A(t)\, I(t)]^T$, system~(\ref{eq:mfSAIR}) can be rewritten in feedback form as:
\begin{align} \label{eq:feedback1}
\dot x(t) & = 
Fx(t) + bu(t){,}\\
\label{eq:feedback2}
y(t) & = c x(t){,}\\
\label{eq:feedback3}
u(t) & = S(t)y(t){,}
\end{align}
where $F$, $b$ and $c$ are defined as
\begin{align} \nonumber
F = \left[ \begin{array}{cc}
- \alpha - \sigma & 0 \\
\alpha & - \mu \\
\end{array} \right], \quad b = \left[ \begin{array}{c}
1 \\
0 \end{array} \right], \quad c = [\bar k_1\gamma \;\, \bar k_2 \lambda].
\end{align}
The remaining variables satisfy the following differential equations:
\begin{align}\label{eq:feedback4}
\dot S(t) & = -S(t)y(t) = -u(t),\\
\label{eq:feedback5}
\dot R(t) & = Ex(t) = [\sigma \; \mu] x(t).
\end{align}

\begin{lemma}\label{lem1}
Given an initial state $z_0 = [S_0\; x_0\; R_0]$, system~(\ref{eq:mfSAIR}) admits the following equilibria: $z^* = (\bar S, 0, \bar R)$, with $\bar S+ \bar R = 1$.
\end{lemma}

\noindent \textit{Proof of Lemma \ref{lem1}}. System~(\ref{eq:mfSAIR}) admits the equilibria given by $(\bar S,0,0,\bar R)$, with $\bar S + \bar R = 1$. This follows from either $S = 0$ or $\bar k_1 \gamma A +\bar k_2 \lambda I = 0$, which in turns means $A = I = 0$ (or both at the same time).

In the first case, if $S = 0$, $\dot A = 0$ and $\dot I = 0$, if and only if $A = 0$ and $I = 0$, and $\dot R = 0$.

In the second case, if $A = I = 0$, then $\dot A = \dot I = 0$ and also $\dot R = 0$. This concludes our proof. \hfill$\blacksquare$\\

A fundamental result on the asymptotic stability of the system in feedback form is given in the following theorem{, where $\mathcal R_0$ is the basic reproduction number defined as the $H_\infty$ norm of the transfer function of system (\ref{eq:feedback1})-(\ref{eq:feedback3}) with constant feedback $\bar S$}.
\begin{theorem}\label{th1}
The feedback system in~(\ref{eq:feedback1})-(\ref{eq:feedback3}) with constant feedback $\bar S$ is asymptotically stable if and only if
\begin{equation} \label{eq:th1}
\bar S < \bar S^* = \frac{1}{\mathcal R_0} = \frac{(\alpha+\sigma)\mu}{\bar k_1\gamma\mu + \bar k_2\lambda\alpha}.
\end{equation}
\end{theorem}

\noindent\textit{Proof of Theorem \ref{th1}}. The transfer function from $u$ to $y$ is $G(s) = c (s\mathbb I_2 - F)^{-1}b$. We can explicitly calculate it as:
\begin{equation}\nonumber
G(s) = [\bar k_1\gamma \;\, \bar k_2\lambda]
\left[ \begin{array}{c}
s+\mu \\
\alpha \end{array} \right]\frac{1}{(s+\alpha+\sigma)(s +\mu)}.
\end{equation}
We can calculate the $H_\infty$ norm of $G(s)$ {as in the following:}
\begin{equation}\nonumber
G(0) = \frac{\bar k_1\gamma\mu + \bar k_2\lambda\alpha}{(\alpha+\sigma)\mu},
\end{equation}
which is equal to the static gain $G_k(0)$ due to the fact that we have a positive system. By standard root locus argument on the positive system for $G_k(s)$, all the roots of the polynomial are in the left-hand plane, i.e. Hurwitz, if and only if
\begin{equation} \nonumber
\bar S < \bar S^* = \frac{(\alpha+\sigma)\mu}{\bar k_1\gamma\mu + \bar k_2\lambda\alpha},
\end{equation}
where $\bar S^* = 1/G(0)$. {In our case, we define the basic reproduction number $\mathcal R_{0}$ as
\begin{equation} \nonumber
\mathcal R_{0} := \frac{1}{\bar S^*} = G(0) = \frac{\bar k_1\gamma\mu + \bar k_2\lambda\alpha}{(\alpha+\sigma)\mu},
\end{equation}
which is the $H_\infty$ norm of the transfer function $G(s)$, and the equilibrium is stable for $\bar S \bar{\mathcal R}_{0} < 1$. Finally, from $\mathcal R_0$, we can also calculate the herd immune threshold (HIT), $P_c$, as:
$$P_c = 1- \frac{1}{\mathcal R_0} = 1 - \bar S^* = 1 - \frac{(\alpha+\sigma)\mu}{\bar k_1\gamma\mu + \bar k_2\lambda\alpha}.$$}
This concludes our proof. \hfill$\blacksquare$\\


\begin{theorem}\label{th2}
For positive initial conditions, the limit values of $\bar S = \lim_{t\rightarrow\infty} S(t)$ and $\bar R = \lim_{t\rightarrow\infty} R(t)$ can be explicitly calculated by
\begin{align}\nonumber
&\log \Bigg(\frac{S(0)}{\bar S}\Bigg) = -cF^{-1}x(0) - \mathcal R_0(S(0) - \bar S),\\
\nonumber
&\bar R = R(0) + EF^{-1}x(0) + EF^{-1}b(S(0) - \bar S).
\end{align}
\end{theorem}

\noindent\textit{Proof of Theorem \ref{th2}}.
From equations~(\ref{eq:feedback1}) and (\ref{eq:feedback4}), we have
\begin{equation} \nonumber
-x(0) = F\int_0^\infty x(\tau)d\tau - b(\bar S - S(0)),
\end{equation}
namely
\begin{equation} \label{eq:intx}
\int_0^\infty x(\tau)d\tau = -F^{-1}x(0) - F^{-1}b (S(0) - \bar S).
\end{equation}
From equation~(\ref{eq:feedback4}), we have
\begin{equation} \nonumber
\log \Big(\frac{S(0)}{\bar S}\Big) = - \log \Big(\frac{\bar S}{S(0)}\Big) = -c \int_0^\infty x(\tau)d\tau.
\end{equation}
Putting together the two equations above yields
\begin{equation} \nonumber
\log \Big(\frac{S(0)}{\bar S}\Big) = -cF^{-1}x(0) - F^{-1}b (S(0) - \bar S).
\end{equation}
The above formula provides that the following inequality holds true: $\bar S \le S(0) \le 1$. Finally, we can calculate $\bar R$ by taking into account equation~(\ref{eq:feedback5}) and the equation~(\ref{eq:intx}) above as in the following:
\begin{equation} \nonumber
\bar R = R(0) + EF^{-1}x(0) + EF^{-1}b(S(0) - \bar S).
\end{equation}
Note that $\mathcal R_0 = -c F^{-1}b$ is the basic reproduction number which is the $H_\infty$ norm of the system in feedback form. This concludes our proof. \hfill$\blacksquare$\\

We now extend our study to assess the presence of a nonzero epidemic threshold for the infection rates in our model. The \emph{epidemic prevalence} is defined as the total number of infected individuals at the end of the epidemic. This threshold can be seen as the limit above which a finite number of individuals get the virus after its spread in the population, namely the epidemic prevalence attains a finite value. On the other hand, if the infection rate is below this threshold, the epidemic prevalence is null\cite{Moreno_2002}. Let us consider system~(\ref{eq:modSAIR}) and, without lack of generality, set the initial conditions $R(0) = 0$ and $S(0) \simeq 1$, which implies that only a very small number of infected individuals $A(0) = I(0) \simeq 0$ is present at the start of the epidemic. The following result provides the value of the epidemic threshold for our case.

\begin{theorem}\label{th3}
Consider system~(\ref{eq:modSAIR}) with initial conditions $R(0) = 0$, $A(0) = I(0) \simeq 0$, $S(0) \simeq 1$. This system admits a nonzero epidemic prevalence if and only if:
\begin{align}\nonumber
\mathcal R_0 < \gamma_c \frac{\bar k_1}{\alpha + \sigma} + \lambda_c  \frac{\bar k_2 \sigma}{(\alpha + \sigma)\mu},
\end{align}
where $\gamma_c$ and $\lambda_c$ are the thresholds for the infection rate of the asymptomatic infected and  symptomatic infected, respectively. These are defined as:
\begin{align}\label{eq:et}
\gamma_c \triangleq \frac{(\alpha + \sigma)}{\bar k_1}(1 - p), \qquad \lambda_c \triangleq \frac{(\alpha + \sigma)\mu}{\bar k_2 \sigma}p,
\end{align}
where $p$ defines the convex space and thus takes values in~$[0 \; 1]$.
\end{theorem}

\noindent\textit{Proof of Theorem \ref{th3}}.
We start by integrating the equation for $S(t)$ in system~(\ref{eq:modSAIR}) as in the following:
\begin{equation} \nonumber
S(t) = S(0)e^{-\int_0^t \phi(\tau)\textrm{d}\tau},
\end{equation}
where the integral is defined as
\begin{equation} \nonumber
\int_0^t \phi(\tau) \textrm{d}\tau = [\bar k_1\gamma \; \bar k_2\lambda] \left[ \begin{array}{cc}
\alpha & -\mu \\
\sigma & \mu \end{array} \right]^{-1} \left[ \begin{array}{c}
I(t) - I(0) \\
R(t) - R(0) \end{array} \right],
\end{equation}
which then yields
\begin{equation} \nonumber
S(t) = S(0)e^{-\big(\frac{\bar k_1 \gamma \mu -  \bar k_2 \lambda \sigma}{(\alpha + \sigma)\mu}(I(t) - I(0)) + \frac{\bar k_1 \gamma \mu + \bar k_2 \lambda \alpha}{(\alpha + \sigma)\mu}(R(t) - R(0))\big)},
\end{equation}
which can be simplified by taking into account the initial conditions, namely $S(0) \simeq 1$ and $I(0) = R(0) = 0$ as specified in the statement of the theorem, and the fact that at the end of the epidemic the number of infected is $\lim_{t\rightarrow \infty}I(t) = 0$ as in the following:
\begin{equation} \nonumber
\bar S = e^{-\mathcal R_0 \bar R},
\end{equation}
where the total number of infected $\bar R = \lim_{t\rightarrow\infty} R(t)$ and $\mathcal R_0$ is the basic reproduction number as defined in Theorem~\ref{th1}.
We can now combine the above equation with the normalization condition and we can see that the total number of infected $\bar R$ fulfils the following equation:
\begin{equation} \nonumber
\bar R = 1 - e^{-\mathcal R_0 \bar R}.
\end{equation}

A trivial solution of the above equation is $\bar R = 0$, but we seek nonzero solutions that are consistent to the stability condition in Theorem~\ref{th1}, for which the following must hold:
\begin{equation} \nonumber
\frac{\textrm{d}}{\textrm{d}\bar R} \Big(1 - e^{-\mathcal R_0 \bar R}\Big)\Big|_{\bar R = 0}<1.
\end{equation}
The above condition is equivalent to the following: 
\begin{align}\nonumber
\mathcal R_0 < \gamma_c \frac{\bar k_1}{\alpha + \sigma} + \lambda_c  \frac{\bar k_2 \sigma}{(\alpha + \sigma)\mu},
\end{align}
which corresponds to the epidemic thresholds in~(\ref{eq:et}). This concludes our proof. \hfill$\blacksquare$\\

Finally, in order to assess the impact of the asymptomatic infected, we now study the dynamics of the ratio between the symptomatic infected and the asymptomatic infected, namely $\tilde I := I/A$. We can calculate the corresponding ODE as:
\begin{align}\label{eq:Iratio}
\dot{\tilde I} = & \frac{\dot I A - I\dot A}{A^2} \\
\nonumber = & \frac{(\alpha A-\mu I)A}{A^2} - \frac{(\bar k_1 \gamma A+\bar k_2\lambda I)IS}{A^2} + \frac{(\alpha+\sigma)IA}{A^2}\\
\nonumber = & \alpha - (\mu +\bar k_1 \gamma S - \alpha - \sigma) \tilde I - \bar k_2 \lambda S \tilde I^2,
\end{align}
and therefore $\dot{\tilde I}$ satisfies a differential Riccati equation as
\begin{align}\label{eq:IRiccati}
\dot{\tilde I} = \alpha - (\mu +\bar k_1 \gamma S - \alpha - \sigma) \tilde I - \bar k_2 \lambda S \tilde I^2,
\end{align}
where the state variables $S$ and $A$ can be rewritten as:
\begin{equation} \label{eq:IratioSA}
\begin{array}{ll}
\dot S = -SA(\bar k_1 \gamma + \bar k_2 \lambda \tilde I), \\
\dot A = SA(\bar k_1 \gamma + \bar k_2 \lambda \tilde I) - A(\alpha+\sigma). \\
\end{array}
\end{equation}
\begin{theorem}\label{th4}
Equation~(\ref{eq:Iratio}) with constant feedback $\bar S$ tends to the equilibrium
\begin{equation} \label{eq:thI}
\bar{\tilde I} = \frac{1}{2 \bar k_2 \lambda \bar S} \Big(h - \bar k_1 \gamma \bar S +\sqrt{(h - \bar k_1 \gamma \bar S)^2 + 4 \alpha  \bar k_2 \lambda \bar S} \Big),
\end{equation}
where $h := \alpha + \sigma - \mu$, and $\bar{\tilde I}$ is asymptotically stable.
\end{theorem}

{\section*{\large Heterogeneous Interaction Model}}\label{sec:CN}
In the previous section, we have studied the model where all individuals in the population are homogeneous, namely {they are indistinguishable}, as they have the same {value to measure the average number of contacts}. In this section, we extend the previous model to address the effects of contact heterogeneity in the form of complex networks. Given a large population, let $P(k)$ be the probability distribution of the node degrees for a complex network representing the interactions of the individuals in the population. Moreover, let  $\langle k \rangle$ be the mean value of the connectivity in the population. 
Let $\theta_i(t) := \frac{1}{\langle f \rangle}\sum_k k P(k) x^{[k]}_i(t)$ be the probability that a randomly chosen link will point to $x^{[k]}_i(t)$ in~(\ref{eq:feedback1}), namely an asymptomatic infected for $i=1$ for any class $k$, and 
a symptomatic infected for $i=2$ for any class $k$, where $\langle f \rangle$ represents the mean value of the connectivity in the population. Finally, let $\psi_{i,k} := k/k_{i,max}${, where $k_{i,max}$ is the maximum number of contacts without restrictions. Parameters $\psi_{i,k}$} describe the connectivity towards the asymptomatic and symptomatic infected for $i=1$ and $i=2$, respectively. {When $k=k_{i,max}$ for all $k$,} we return to the homogeneous case.

Let $z_k(t) = [S_k(t)\, A_k(t)\, I_k(t)\, R_k(t)]^T$ be the population state at time $t$ of connectivity $k$. 
The magnitudes $S_k(t)$, $A_k(t)$, $I_k(t)$ and $R_k(t)$ represent the density of the susceptible, asymptomatic infected, symptomatic infected and removed nodes of connectivity $k$ at time $t$, respectively. As before, these variables must satisfy the normalization condition for each $k \in \mathbb Z$:
$$S_k(t) + A_k(t) + I_k(t) + R_k(t) = 1.$$

For each population $k \in \mathbb Z$, system (\ref{eq:modSAIR}) becomes: 
\begin{equation} \label{eq:modCN}
\left\{\begin{array}{ll}
\dot S_k(t) = -S_k(t)(\psi_{1,k} \gamma \theta_1(t) + \psi_{2,k} \lambda \theta_2(t)), \\
\dot A_k(t) = S_k(t)(\psi_{1,k} \gamma \theta_1(t) + \psi_{2,k} \lambda \theta_2(t)) - A_k(t)(\alpha + \sigma), \\
\dot I_k(t) = \alpha A_k(t)  - \mu I_k(t), \\
\dot R_k(t) = \sigma A_k(t) + \mu I_k(t). \\
\end{array} \right.
\end{equation}

Each node of the network represents an individual and their corresponding state, i.e. susceptible, asymptomatic infected, symptomatic infected and removed. In matrix form, where the dependence on time is implicit for {the sake of} brevity, the above system becomes:
\begin{equation} \label{eq:mfSAIRCN}
\begin{scriptsize}
\left[ \begin{array}{c}
\dot S_k \\
\dot A_k \\
\dot I_k \\
\dot R_k \end{array} \right] = 
\underbrace{\left[ \begin{array}{cccc}
- (\psi_{1,k} \gamma \theta_1 + \psi_{2,k} \lambda \theta_2) & 0 & 0 & 0 \\
(\psi_{1,k} \gamma \theta_1 + \psi_{2,k} \lambda \theta_2) &  - (\alpha + \sigma) & 0 & 0 \\
0 & \alpha & - \mu & 0 \\
0 & \sigma & \mu & 0 \\ 
\end{array} \right]}_{G_k(\theta)} 
\left[ \begin{array}{c}
S_k \\
A_k \\
I_k \\
R_k \end{array} \right],
\end{scriptsize}
\end{equation}
where $\theta := [\theta_1 \, \theta_2]^T$ and $G_k(\theta)$ depends explicitly on the measure of connectivity $k$ and on $\theta$.

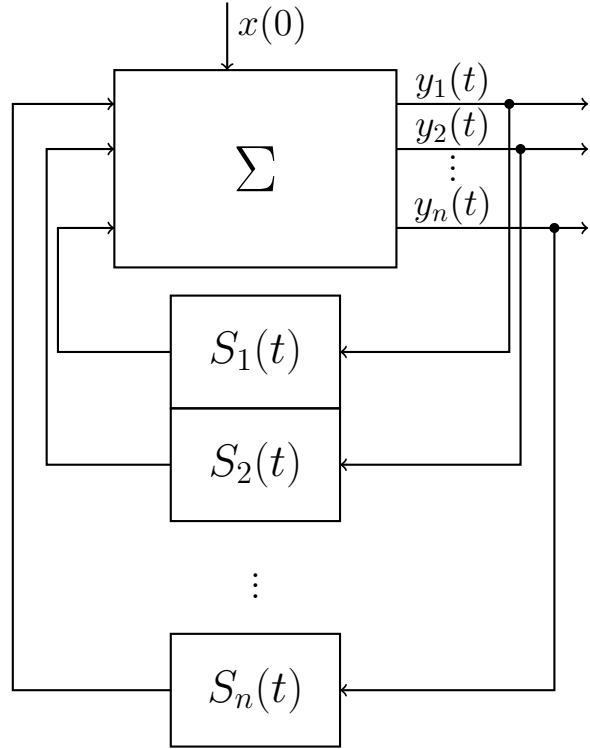
\begin{figure}[t]
\centering
	\begin{tikzpicture}
	\begin{scope}[shift={(-1,0)},scale=1.5]
		\draw [thick] (-.5,-.75) rectangle (2,1);
		\node at (.75,0.125) {\Huge $\Sigma$};
		
		\draw [thick] (0,-2) rectangle (1.5,-1);
		\node at (.75,-1.5) {\Large $S_1(t)$};
		\draw [thick] (0,-3) rectangle (1.5,-2);
		\node at (.75,-2.5) {\Large $S_2(t)$};
		\node at (.75,-3.5) {\Large $\vdots$};
		\draw [thick] (0,-5) rectangle (1.5,-4);
		\node at (.75,-4.5) {\Large $S_n(t)$};
		
		\draw [->,  thick]  (0,-1.5) -- (-1,-1.5) -- (-1,-.4) -- (-0.5,-.4);
		\draw [->,  thick]  (0,-2.5) -- (-1.1,-2.5) -- (-1.1,.3) -- (-0.5,.3);
		\draw [->,  thick]  (0,-4.5) -- (-1.4,-4.5) -- (-1.4,.7) -- (-0.5,.7);
		
		\draw [->,  thick] (0.5,1.6) -- (0.5,1);
		\node at (.9,1.4) {\large $x(0)$};
		
		\draw [->, thick] (3,.7) -- (3,-1.5) -- (1.5,-1.5);
		\draw [thick,fill=black] (3,.7) circle (1pt);
		\draw [->, thick] (3.1,.3) -- (3.1,-2.5) -- (1.5,-2.5);
		\draw [thick,fill=black] (3.1,.3) circle (1pt);
		\draw [->, thick] (3.4,-.4) -- (3.4,-4.5) -- (1.5,-4.5);
		\draw [thick,fill=black] (3.4,-.4) circle (1pt);
		
		
		\draw [->, thick] (2,.7) -- (3.7,.7);
		\node at (2.5,.9) {\large $y_1(t)$};
		\draw [->, thick] (2,.3) -- (3.7,.3);
		\node at (2.5,.5) {\large $y_2(t)$};
		\node at (2.5,.2) {\large $\vdots$};
		\draw [->, thick] (2,-.4) -- (3.7,-.4);
		\node at (2.5,-.2) {\large $y_n(t)$};
	\end{scope}
	\end{tikzpicture}
	\caption{{The heterogeneous SAIR system in feedback form corresponding to equations in~(\ref{eq:feedbackCN1})-(\ref{eq:feedbackCN3}).}} 
\label{fig:CNfeedback}
\end{figure}

As for the homogeneous case, we can rewrite the above system in feedback form. We start by writing each system corresponding to connectivity $k$ and then we write the whole system comprising all $k \in \mathbb Z$, for $k \in [1 \; N]$, where $N$ is the highest connectivity. Let $x^{[k]}(t) = [A_k(t)\, I_k(t)]^T$, system~(\ref{eq:mfSAIRCN}) in feedback form is the following:
\begin{align}  \label{eq:feedbackCN1}
\dot x^{[k]}(t) & = 
Fx^{[k]}(t) + bu_k(t),\\
\label{eq:feedbackCN2}
y_k(t) & = c_k \sum_j jP(j)x^{[j]}(t), \\
\label{eq:feedbackCN3}
u_k(t) & = S_k(t)y_k(t),
\end{align}
where $F$ and $b$ are defined as in the homogeneous case and $c_k$ is:
\begin{align} \nonumber 
c_k = \Bigg[\frac{\psi_{1,k}\gamma}{\langle f \rangle} \;\, \frac{\psi_{2,k} \lambda}{\langle f \rangle}\Bigg],
\end{align}
where $x^{[j]}(t) = [A_j(t) \; I_j(t)]^T$ with the dummy variable $j$ for the sum.
The remaining variables satisfy the following differential equations:
\begin{align}\label{eq:feedbackCN4}
\dot S_k(t) & = -S_k(t)y_k(t) = -u_k(t),\\
\label{eq:feedbackCN5}
\dot R_k(t) & = Ex^{[k]}(t) = [\sigma \; \mu] x^{[k]}(t).
\end{align}
The system consisting of all subsystems for all $k$ is the following:
\begin{equation} \label{eq:mfSAIRCNx}
\dot x(t) = (\mathbb I_N \otimes F)x + \textrm{diag}(b) \textrm{diag}(\bar S) c Px,
\end{equation}
where the system is with constant feedback $\bar S$, $c = [c_1^T \; c_2^T \; \cdots \; c_n^T]^T$ and $P = [P(1)\mathbb I_2 \; 2P(2)\mathbb I_2  \cdots nP(n)\mathbb I_2]$, where $\mathbb I_N$ is the $N\times N$ identity matrix, diag$(a)$ is the matrix whose diagonal is $a$ for any vector $a$, and $A \otimes B$ is the Kronecker product between matrix $A$ and matrix $B$. Matrix $A_{BIG}$ for the whole system is the following:
\begin{equation} \label{eq:mfSAIRCNA}
A_{BIG} = \mathbb I_N \otimes F + \textrm{diag}(b) \textrm{diag}(\bar S) c P.
\end{equation}

We are now ready to establish the next result, which assesses the stability of each subsystem with connectivity $k$ in feedback form in~(\ref{eq:feedbackCN1})-(\ref{eq:feedbackCN3}).
\begin{theorem}\label{th5}
The subsystem for class of connectivity $k$ in~(\ref{eq:feedbackCN1})-(\ref{eq:feedbackCN3}) with constant feedback $\bar S_k$ is asymptotically stable if and only if
\begin{equation} \label{eq:th5}
{\bar S_k < \bar S_k^* = \frac{1}{\mathcal R_{0,k}} = \frac{\langle f \rangle(\alpha + \sigma)\mu}{\psi_{1,k}\gamma\mu + \psi_{2,k}\lambda\alpha}.}
\end{equation}
\end{theorem}

\noindent{\textit{Proof of Theorem \ref{th5}}. The transfer function from $u_k$ to $y_k$ for each subsystem for class of connectivity $k$ is $G_k(s) = c_k (s\mathbb I_2 - F)^{-1}b$. We can explicitly calculate it as:
\begin{equation}\nonumber
G_k(s) = \Bigg[\frac{\psi_{1,k}\gamma}{\langle f \rangle} \;\, \frac{\psi_{2,k} \lambda}{\langle f \rangle}\Bigg]
\left[ \begin{array}{c}
s+\mu \\
\alpha \end{array} \right]\frac{1}{(s+\alpha+\sigma)(s +\mu)}.
\end{equation}
We can calculate the $H_\infty$ norm of $G_k(s)$ which is equal to the static gain $G_k(0)$ as we are in the presence of a positive system as in the following:
\begin{equation}\nonumber
G_k(0) = \frac{\psi_{1,k}\gamma\mu + \psi_{2,k}\lambda\alpha}{\langle f \rangle(\alpha+\sigma)\mu}.
\end{equation}
As for the original model, we use the standard root locus argument on the positive system $G_k(s)$, and all the roots of the polynomial are in the left-hand plane, i.e. Hurwitz, if and only if
\begin{equation} \nonumber
\bar S_k < \bar S_k^* = \frac{\langle f \rangle(\alpha + \sigma)\mu}{\psi_{1,k}\gamma\mu + \psi_{2,k}\lambda\alpha},
\end{equation}
where $\bar S_k^* = 1/G_k(0)$ as defined before. Furthermore, we define the basic reproduction number $\mathcal R_{0,k}$ for class of connectivity $k$ as
\begin{equation} \nonumber
\mathcal R_{0,k} := \frac{1}{\bar S_k^*} = G_k(0) = \frac{\psi_{1,k}\gamma\mu + \psi_{2,k}\lambda\alpha}{\langle f \rangle(\alpha + \sigma)\mu},
\end{equation}
which is the $H_\infty$ norm of the transfer function $G_k(s)$, and the equilibrium is stable for $\bar S_k \bar{\mathcal R}_{0,k} < 1$. This concludes our proof. \hfill$\blacksquare$}\\

The next theorem extends the previous result to the system~(\ref{eq:mfSAIRCNx}), consisting of all classes of connectivity.
\begin{theorem}\label{th6}
System~(\ref{eq:mfSAIRCNx}) with constant feedback $\textrm{diag}(\bar S)$ is asymptotically stable if and only if 
$$\sum_j j P(j) \bar S_j \mathcal R_{0,j} < 1,$$
where the basic reproduction number for the whole network $\mathcal R_{0,net}$ is defined as:
\begin{equation} \label{eq:th6}
\mathcal R_{0,\textrm{net}} = \sum_j j P(j) \frac{\psi_{1,j}\gamma\mu+\psi_{2,j}\lambda\alpha}{(\alpha + \sigma)\mu}.
\end{equation}
\end{theorem}

\noindent{\textit{Proof of Theorem \ref{th6}}. The condition $\mathcal R_{0,net} \textrm{diag}(\bar S) < \mathbb I$ is equivalent to saying that the largest eigenvalue of system~(\ref{eq:mfSAIRCNx}) must be less than one, namely $\mathcal R_{0,k} < 1, \; \forall k$. We calculate
\begin{equation} \nonumber
\det(s\mathbb I - A_{BIG}) = 0,
\end{equation}
from which we have
\begin{equation} \nonumber
\det(\mathbb I_2 - P(s\mathbb I_{2N} - \textrm{diag}(F))^{-1}\textrm{diag}(b)\textrm{diag}(\bar S)c) = 0.
\end{equation}
The above equation can be rewritten as
\begin{equation} \nonumber
\det\Bigg(\mathbb I_2 - \frac{\sum_i i P(i) \bar S_i \left[ \begin{array}{c}
s + \mu \\
\alpha \end{array} \right] c_i}{(s-\mu)(s - \alpha - \sigma)} \Bigg)= 0,
\end{equation}
where we can now set $s = 0$ to get to the following:
\begin{equation} \nonumber
\mu(\alpha + \sigma) > \sum_i i P(i) \bar S_i [c_{1,i}\mu + c_{2,i} \alpha].
\end{equation}
From the above, we find the formula in (\ref{eq:th6}). This concludes our proof. \hfill$\blacksquare$}\\

Similarly to the homogeneous case, we provide a calculation of the nonzero epidemic threshold in the case of structured environment. Without loss of generality, let us consider system~(\ref{eq:modCN}) with the following initial conditions, identical for all classes $k$: $R_k(0) = 0$ and $S_k(0) \simeq 1$, for which $A_k(0) = I_k(0) \simeq 0$. We find an expression for the epidemic threshold in the case of complex networks in the following result.

\begin{theorem}\label{th7}
Consider system~(\ref{eq:modCN}) with initial conditions $R_k(0) = 0$, $A_k(0) = I_k(0) \simeq 0$, $S_k(0) \simeq 1$. This system admits a nonzero epidemic prevalence if and only if:
\begin{align}\nonumber
\mathcal R_{0,k} < \gamma_c \frac{\psi_{1,k}}{\langle f \rangle(\alpha + \sigma)} + \lambda_c  \frac{\psi_{2,k} \sigma}{\langle f \rangle(\alpha + \sigma)\mu},
\end{align}
where $\gamma_c$ and $\lambda_c$ are the thresholds for the structured case and are defined as in the following:
\begin{align}\label{eq:etCN}
\gamma_c \triangleq \frac{\langle f \rangle(\alpha + \sigma)}{\psi_{1,k}}(1 - p), \qquad \lambda_c \triangleq \frac{\langle f \rangle(\alpha + \sigma)\mu}{\psi_{2,k} \sigma}p,
\end{align}
where $p$ defines the convex space and thus takes values in~$[0 \; 1]$.
\end{theorem}

\noindent\textit{Proof of Theorem \ref{th7}}.
Consider the equation for $S_k(t)$ in system~(\ref{eq:modCN}), and let us integrate it as in the following:
\begin{equation} \nonumber
S_k(t) = S_k(0)e^{-\int_0^t \phi_k(\tau)\textrm{d}\tau},
\end{equation}
where the integral is defined as
\begin{align} \nonumber
\int_0^t \phi_k(\tau) \textrm{d}\tau = \Bigg[\frac{\psi_{1,k}\gamma}{\langle f \rangle} \; \frac{\psi_{2,k}\lambda}{\langle f \rangle}\Bigg] & \left[ \begin{array}{cc}
\alpha & -\mu \\
\sigma & \mu \end{array} \right]^{-1} \\
& \left[ \begin{array}{c}
I_k(t) - I_k(0) \\
R_k(t) - R_k(0) \end{array} \right],
\end{align}
which we can calculate as
\begin{align} \nonumber
&S_k(t) = \\
&S_k(0)e^{-\big(\frac{\psi_{1,k} \gamma \mu -  \psi_{2,k} \lambda \sigma}{\langle f \rangle(\alpha + \sigma)\mu}(I_k(t) - I_k(0)) + \frac{\psi_{1,k} \gamma \mu + \psi_{2,k} \lambda \alpha}{\langle f \rangle(\alpha + \sigma)\mu}(R_k(t) - R_k(0))\big)}.
\end{align}

By taking into account the initial conditions, namely $S_k(0) \simeq 1$ and $I_k(0) = R_k(0) = 0$, and the fact that at the end of the epidemic the number of infected is $\lim_{t\rightarrow \infty}I_k(t) = 0$, the following holds:
\begin{equation} \nonumber
\bar S = e^{-\mathcal R_{0,k} \bar R_k},
\end{equation}
where the total number of infected for each class $k$ is $\bar R_k = \lim_{t\rightarrow\infty} R(t)$ and $\mathcal R_{0,k}$ is the basic reproduction number from Theorem~\ref{th5}. We can now combine the above equation with the normalization condition and we can see that the total number of infected $\bar R_k$ fulfils the following equation:
\begin{equation} \nonumber
\bar R_k = 1 - e^{-\mathcal R_{,k}0 \bar R_k}.
\end{equation}

As before, we can find a trivial solution of the above equation for $\bar R_k = 0$, but for nonzero solutions that are consistent to the stability condition in Theorem~\ref{th5}, the following must hold:
\begin{equation} \nonumber
\frac{\textrm{d}}{\textrm{d}\bar R_k} \Big(1 - e^{-\mathcal R_{0,k} \bar R_k}\Big)\Big|_{\bar R_k = 0}<1.
\end{equation}
The above condition is equivalent to the following: 
\begin{align}\nonumber
\mathcal R_{0,k} < \gamma_c \frac{\psi_{1,k}}{\langle f \rangle(\alpha + \sigma)} + \lambda_c  \frac{\psi_{2,k} \sigma}{\langle f \rangle(\alpha + \sigma)\mu},
\end{align}
which corresponds to the epidemic thresholds in~(\ref{eq:etCN}).
This concludes our proof. \hfill$\blacksquare$\\

\begin{figure}[h!]
    \centering
    \includegraphics[width=\textwidth]{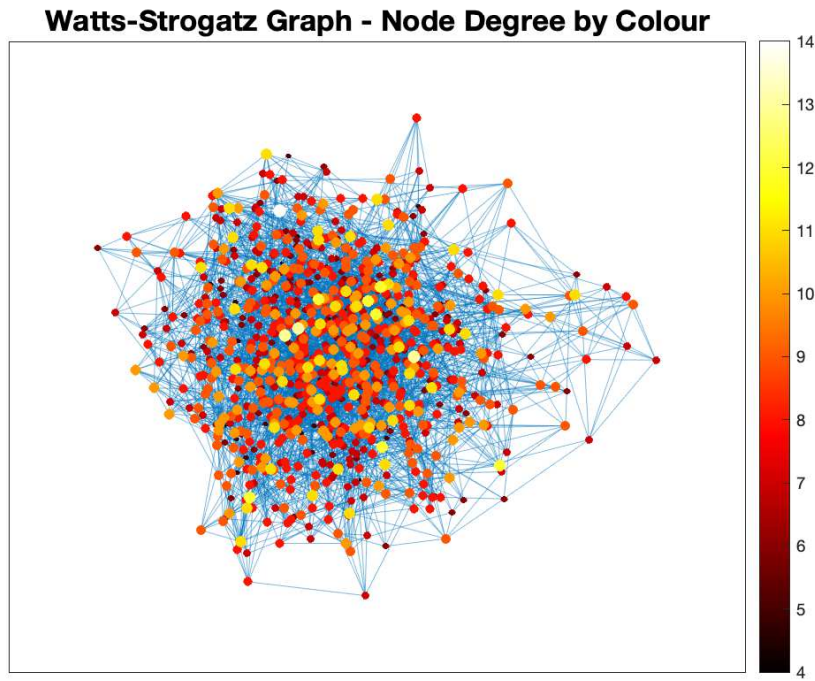}
    \caption{{Small world network with $N=1000$, $m=4$ and $p=1$, where the colour of each node corresponds to its node degree as in the colorbar.}}
    \label{fig:CN}
\end{figure}

\begin{figure}[h!]
    \centering
    \includegraphics[width=\textwidth]{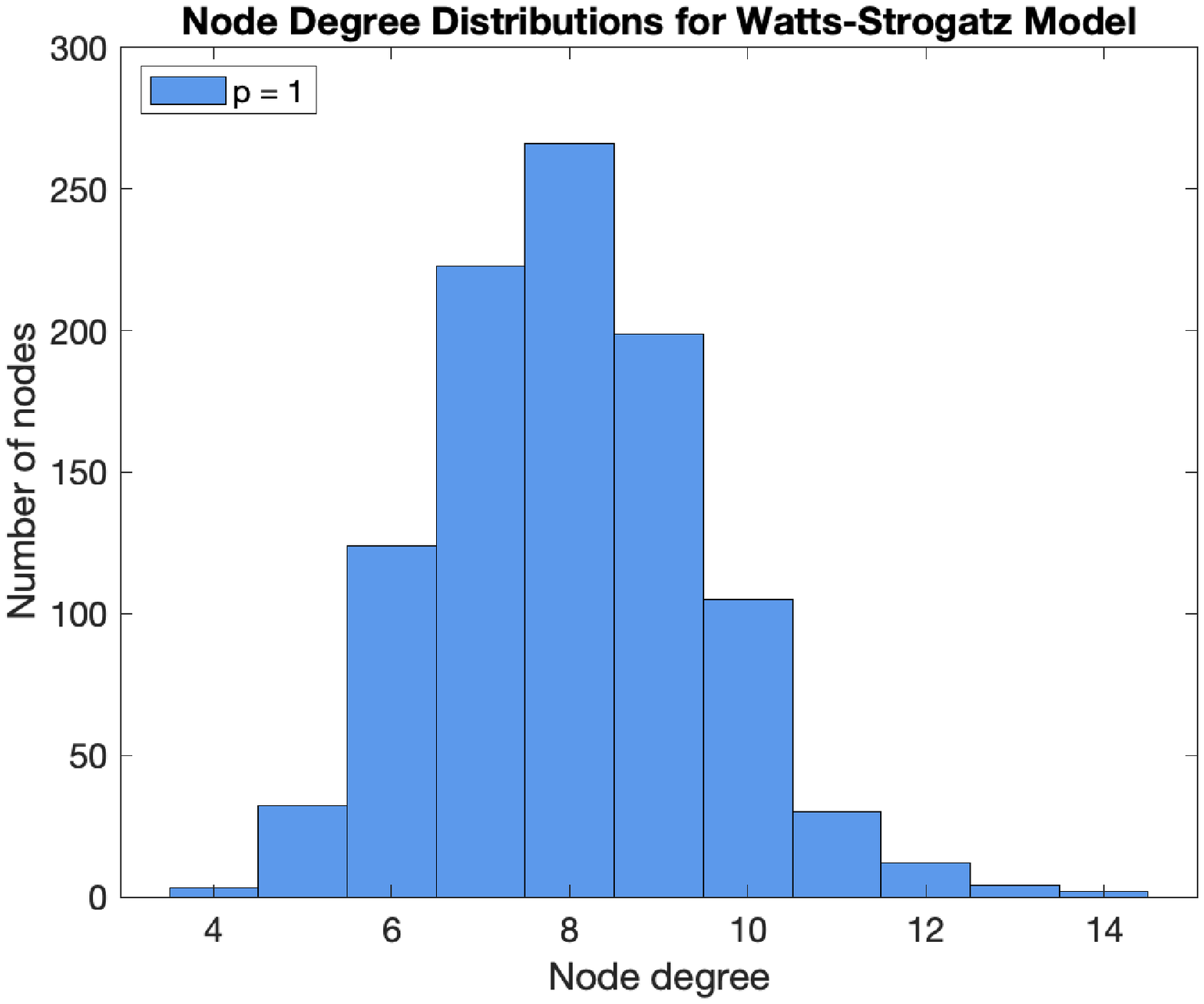}
    \caption{{Small world network: distribution of node degree for $N=1000$, $m=4$ and $p=1$.}}
    \label{fig:CN-ND}
\end{figure}

{\section*{\large Numerical Analysis:  Small World Network}}
In this section, we present the numerical analysis conducted on the Watts-Strogatz model to show the impact of heterogeneous connectivity in system~(\ref{eq:modCN}). For the purpose of illustration, we consider a WS model for $N=1000$ nodes, given $\langle k \rangle = 2m$ and $m = 4$. To generate the network we use a discretised version of the following formula
\begin{equation} \label{eq:Pk}
P(k) = \frac{m^{(k-m)}}{(k-m)!e^{-m}} \qquad  \textrm{for} \; k \ge m,
\end{equation}
where the node degrees vary between 4 and 14. The discretised version is obtained from discarding the values less than 4 and greater than 14, and rounding up the fractions of the populations in the other classes such that the total population across the classes sums up to 1. We also set $p=1$, where $p$ is the probability of rewiring a node from the starting ring graph, each node being connected to its $2m$ nearest neighbours\cite{Barrat_2000}. Figure~\ref{fig:CN} shows the corresponding WS complex network, where each node has a colour corresponding to its node degree as in the colorbar on the right of the figure. Figure~\ref{fig:CN-ND} shows the histogram of the node degrees for all nodes in the network.

\begin{figure}[h!]
    \centering
    \includegraphics[width=\textwidth]{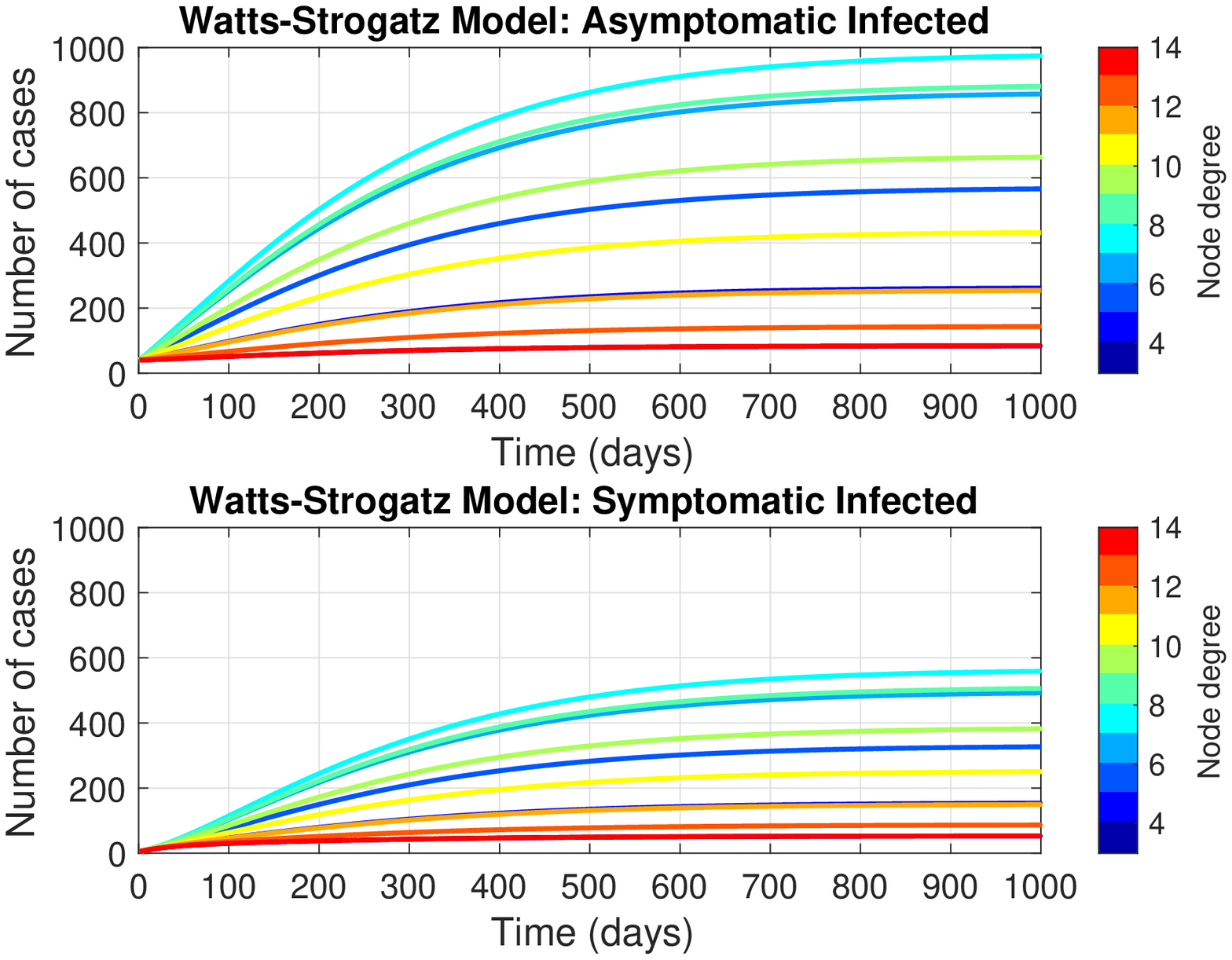}
    \caption{{WS Model: time evolution of total cumulative asymptomatic (top) and symptomatic infected (bottom).}}
    \label{fig:CN-NA-AI}
\end{figure}

\begin{figure}[h!]
    \centering
    \includegraphics[width=\textwidth]{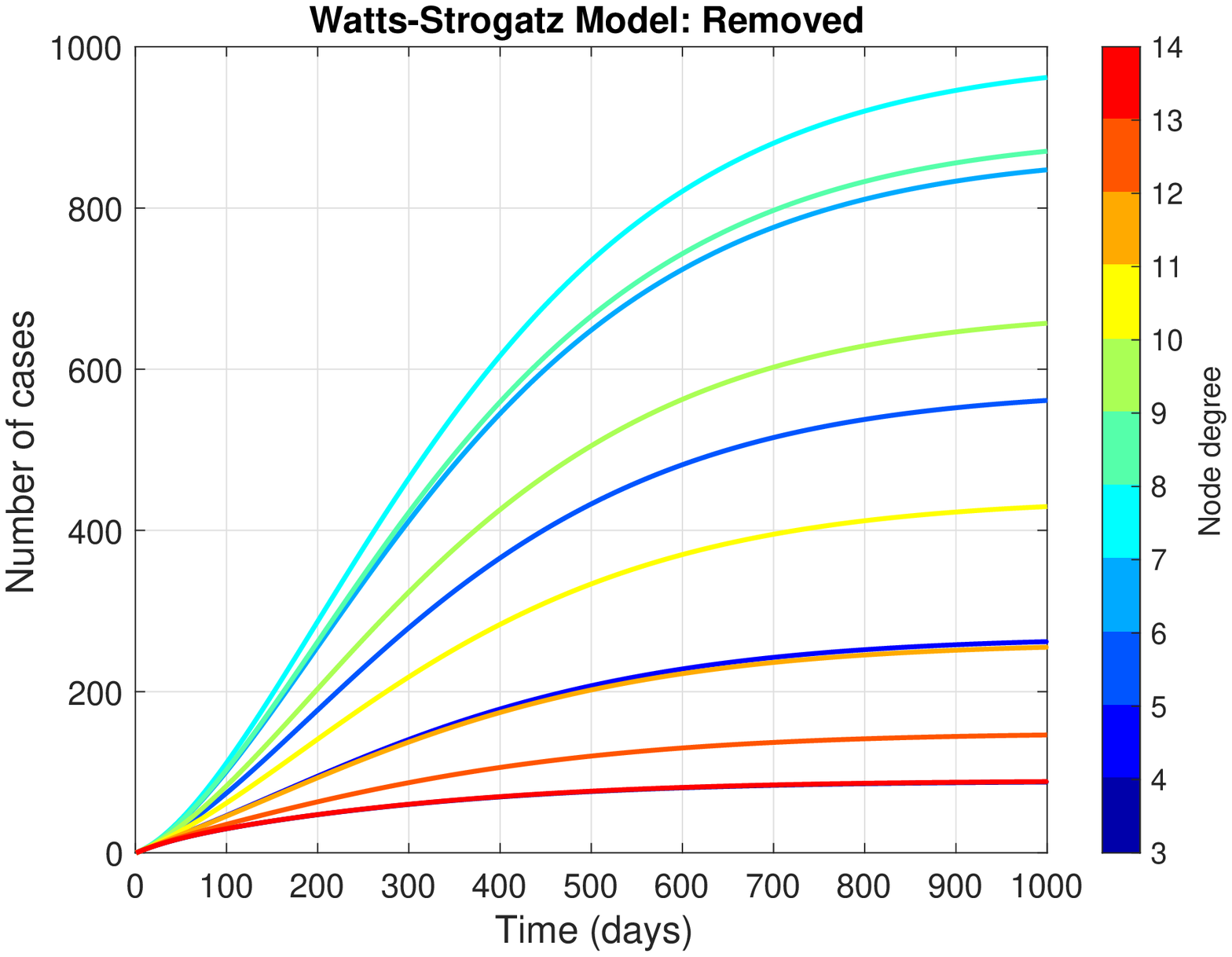}
    \caption{{WS Model: time evolution of removed.}}
    \label{fig:CN-NA-R}
\end{figure}

For the simulations, we consider a WS complex network with $N = 10^7$ nodes, $\langle k \rangle = 2m = 8$ and $k_{2,max}/k_{1,max} = 0.4$. The parameters for each class of connectivity are shown in Table~\ref{t:data}, whilst the constant parameters are set as in the following: $\gamma=0.447$, $\sigma=0.087$, $\lambda=0.469$, $\mu=0.0523$, and $\alpha=0.115$. 

\begin{table} [h!]
\begin{center}
   	\caption{Parameters for each class of connectivity.}
  	\begin{tabular}{|c|c|c|c|c|}
   	\hline
   	$k$ & Distribution & $\psi_{1_k}$ & $\psi_{2_k}$ \\
   	\hline \hline
   	$4$ & $0.0183$ & $0.2000$ & $0.0800$ \\
   	\hline
   	$5$ & $0.0741$ & $0.2500$ & $0.1000$ \\
   	\hline
	$6$ & $0.1469$ & $0.3000$ & $0.1200$ \\
   	\hline
	$7$ & $0.1957$ & $0.3500$ & $0.1400$ \\
   	\hline
	$8$ & $0.1957$ & $0.4000$ & $0.1600$ \\
   	\hline
	$9$ & $0.1566$ & $0.4500$ & $0.1800$ \\
   	\hline
	$10$ & $0.1045$ & $0.5000$ & $0.2000$ \\
   	\hline
	$11$ & $0.0597$ & $0.5500$ & $0.2200$ \\
   	\hline
	$12$ & $0.0299$ & $0.6000$ & $0.2400$ \\
   	\hline
	$13$ & $0.0133$ & $0.6500$ & $0.2600$ \\
   	\hline
	$14$ & $0.0053$ & $0.7000$ & $0.2800$ \\
   	\hline
   	\end{tabular}
	\label{t:data}
\end{center}
\end{table}    

Finally, we update $\psi_{1_k}$ and $\psi_{2_k}$ by decreasing them by 0.005 and 0.00135 of their initial value after $t>10$. Figures~\ref{fig:CN-NA-AI}-\ref{fig:CN-NA-R} depict the propagation of the disease for each class in terms of the asymptomatic and symptomatic infected first, namely $A_k(t)$ and $I_k(t)$, and then in terms of the removed, i.e. $R_k(t)$. Each class is shown in a different colour corresponding to the colorbar on the right hand side.

{\section*{\LARGE Results}}\label{sec:model}\label{sec:CS}
In this section, we validate our models with the official data from Dipartimento della Protezione Civile\cite{Protezione_Civile,pcm_dpc}, and also we provide an investigation on the impact of asymptomatic infected through the recent seroprevalence study conducted by Istat\cite{Istat_2020}. We provide two case studies, the first one via the homogeneous model and the second one via complex networks. The first case study includes two sets of simulations: in the first one, we use the official data to tune the parameters of our model and we compare this with the total cumulative infected and with the estimated number of individuals with antibodies found in the seroprevalence study; in the second one, we do the opposite, i.e. we fit our model with the seroprevalence study and compare our model to the official data. In the second case study, we investigate the interactions among individuals from different regions in Italy, first. Finally, we provide a prediction on the evolution of the pandemic for a specific region (Campania), given different containment strategies in the context of school opening.


\begin{figure}[t]
    \centering
    \includegraphics[width=\textwidth]{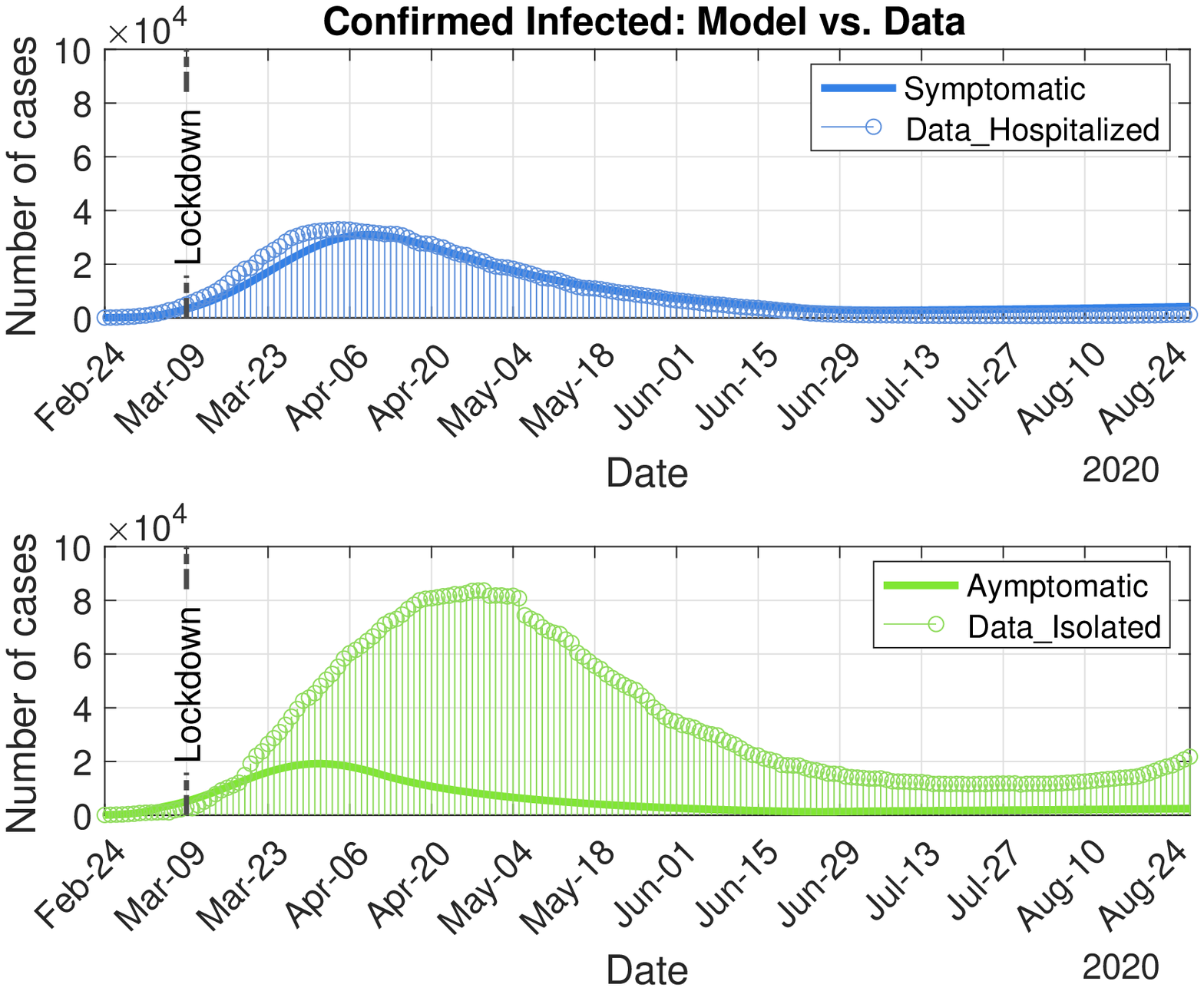}
    \caption{Confirmed infected: model vs. data, symptomatic and asymptomatic against hospitalised and isolated\cite{pcm_dpc}.}
    \label{fig:italy_data_nosero}
\end{figure}

\begin{figure}[t]
    \centering
    \includegraphics[width=\textwidth]{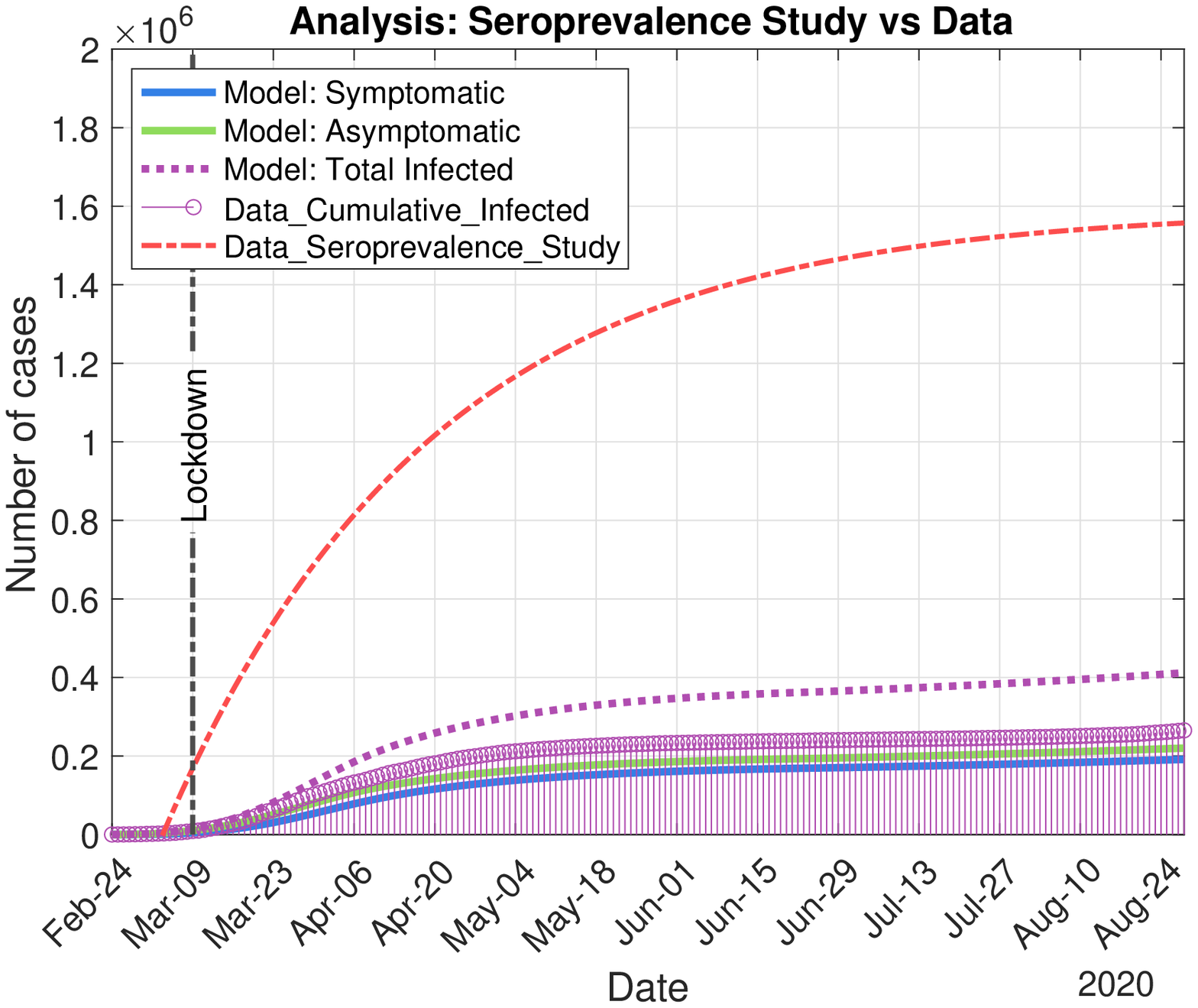}
    \caption{Analysis: seroprevalence study vs data\cite{Istat_2020}.}
    \label{fig:italySeroprevalence_nosero}
\end{figure}

\subsection*{Homogeneous Model: Data and Seroprevalence Study}
In the first investigation, we use the official data to fit our model and estimate the parameters and then we compare our model to the value of the Istat seroprevalence study. We set the portion of the population in each stage as: $A(0) = 94/(60*10^6)$, $I(0) = 127/(60*10^6)$, $R(0) = 0$, and $S(0) = 1-A(0)-I(0)-R(0)$, where these values are taken from the data for the isolated at home and hospitalised infected\cite{pcm_dpc}. The reason behind this choice is that we believe that people that are not hospitalised must either be asymptomatic or paucisymptomatic and thus would fall in our category of asymptomatic infected. 
The other parameters are set as in the following:
\begin{eqnarray*}
\gamma&=&0.46952, \; \sigma=0.025501, \; \bar k_1 = 0.99209, \\
\lambda&=&0.48521, \; \mu=0.10004, \;  \bar k_2 = 0.65056,\\
\alpha&=&0.185017.
\end{eqnarray*}

{To have an initial estimate of the parameters, we used an implementation of a non-linear least squares regression, using the Levenberg-Marquardt algorithm\cite{Griva_2008}. This is an iterative optimisation algorithm that fits a function to a desired output, obtaining the parameters that minimises the square error between the output of the function and the objective value given. In this specific case, the values that were fit were the number of symptomatic active cases and the number of removed. This algorithm is widely used because of its versatility and efficient use of data, even on small datasets\cite{NIST_2003}. However, it is very sensitive to the hyperparameters so an educated initial estimation of them was provided as well as the specific range of values that each parameter could take. These parameter values, which were analytically extracted, were used as a starting point, and were later adapted to better match the empirical results.}
 Due to the similar viral load between symptomatic and asymptomatic individuals\cite{Zou_2020}, we set the values of $\gamma$ and $\lambda$ to be very close. Parameter $\bar k_1$ is chosen to be larger than $\bar k_2$ at the start (and also in future time instants), because it accounts for the likelihood that people interact with asymptomatic individuals more likely than with infected that show symptoms. On March 6th, prime minister Giuseppe Conte imposed a set of localised lockdowns to isolate the outbreaks, and on March 9th a national quarantine was imposed, which restricted the movements of the population and therefore their contacts and interactions. We account for this by lowering the values of $\bar k_1$ and $\bar k_2$ slowly over the days following the lockdown, down to $\bar k_1 = 0.2957$ and $\bar k_2 = 0.0305$ before the end of the quarantine period. Following the ease of the lockdown measures, we set $\bar k_1 = 0.3636$ and $\bar k_2 = 0.0594$ to account for the increased interactions during mid-August holidays. At the end of February and thus before the lockdown, we calculate $\mathcal R_0 = 4.98$, in accordance with studies that place it between 2 and 5\cite{Wu_2020, Zhao_2020, Anastassopoulou_2020, Gatto_2020}. Towards the end of the quarantine, the value of $\mathcal R_0$ goes below 1 and then it oscillates around $\mathcal R_0 = 1.06$ during August. As it can be seen in Fig.~\ref{fig:italy_data_nosero}, our model matches quite accurately the recovered and hospitalised infected, but it does not do the same with the asymptomatic infected. Even in that case, we can see from Fig.~\ref{fig:italySeroprevalence_nosero} that our estimation of the cumulative infected is higher that the confirmed cases. It is matching quite closely an early estimate of the undetected asymptomatic being around $30\%$, but far away from the current Istat estimate depicted in red. In accordance with Theorem~\ref{th3}, the ratio $\bar{\tilde I} = 1.7041$, which is the exact ratio between $\bar I$ and $\bar A$. {When using the work of B\"ohning et al. to estimate the hidden infection, we obtain almost identical values, namely a number of $264240$ hidden infected\cite{Bohning_2020}. This value would account for twice as many infected individuals as the number of detected infections, but it is underestimated if compared with seroprevalence studies\cite{Istat_2020,Pollan_2020}.}

\begin{figure}[h!]
        \centering
        \begin{subfigure}[b]{.7\textwidth}
            \includegraphics[width=\textwidth]{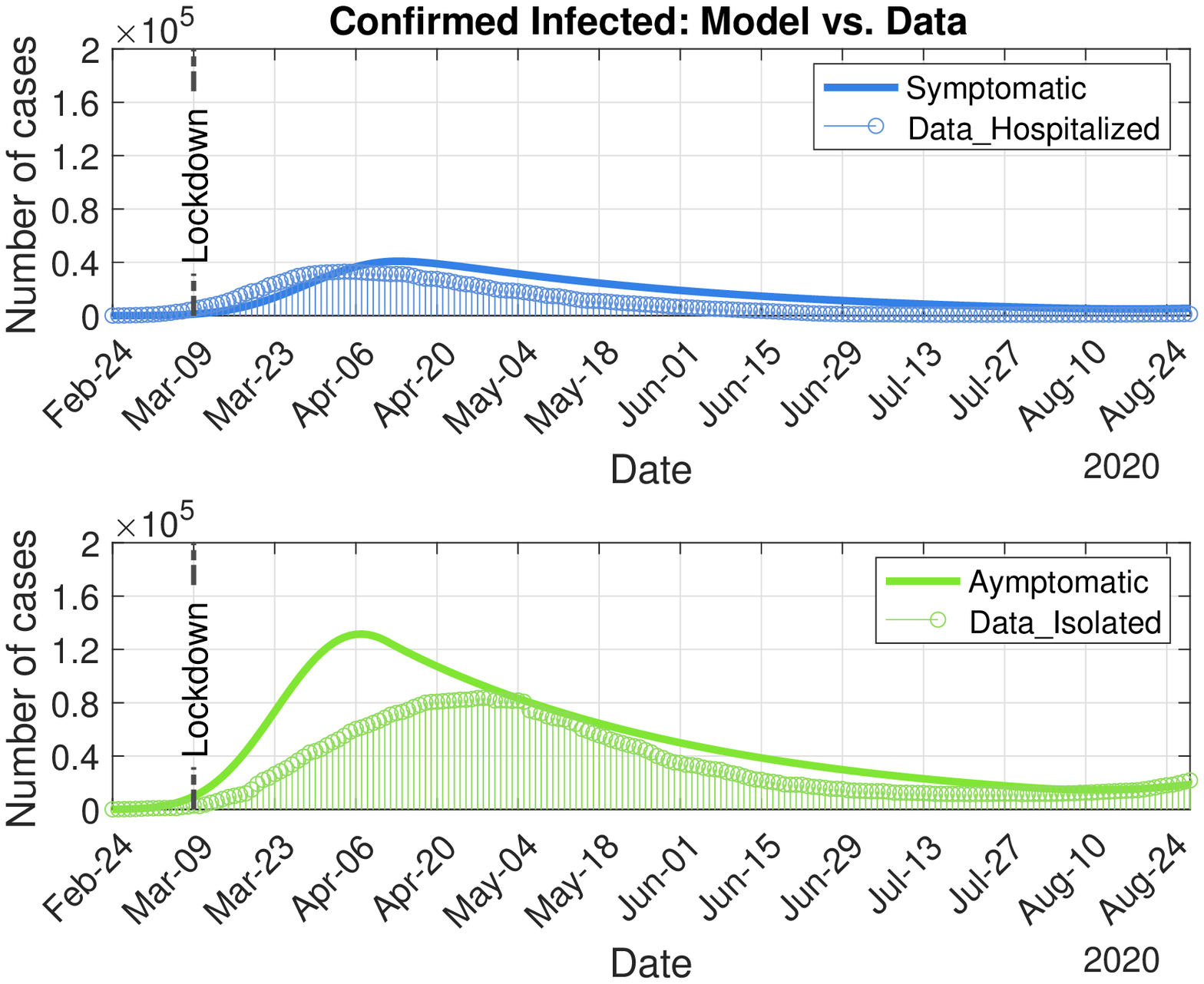}
            \caption{{\small Confirmed infected: model vs. data\cite{pcm_dpc}.}}    
            \label{fig:italyAS_data2}
        \end{subfigure}
        \vskip\baselineskip
        \begin{subfigure}[b]{.7\textwidth}   
            \includegraphics[width=\textwidth]{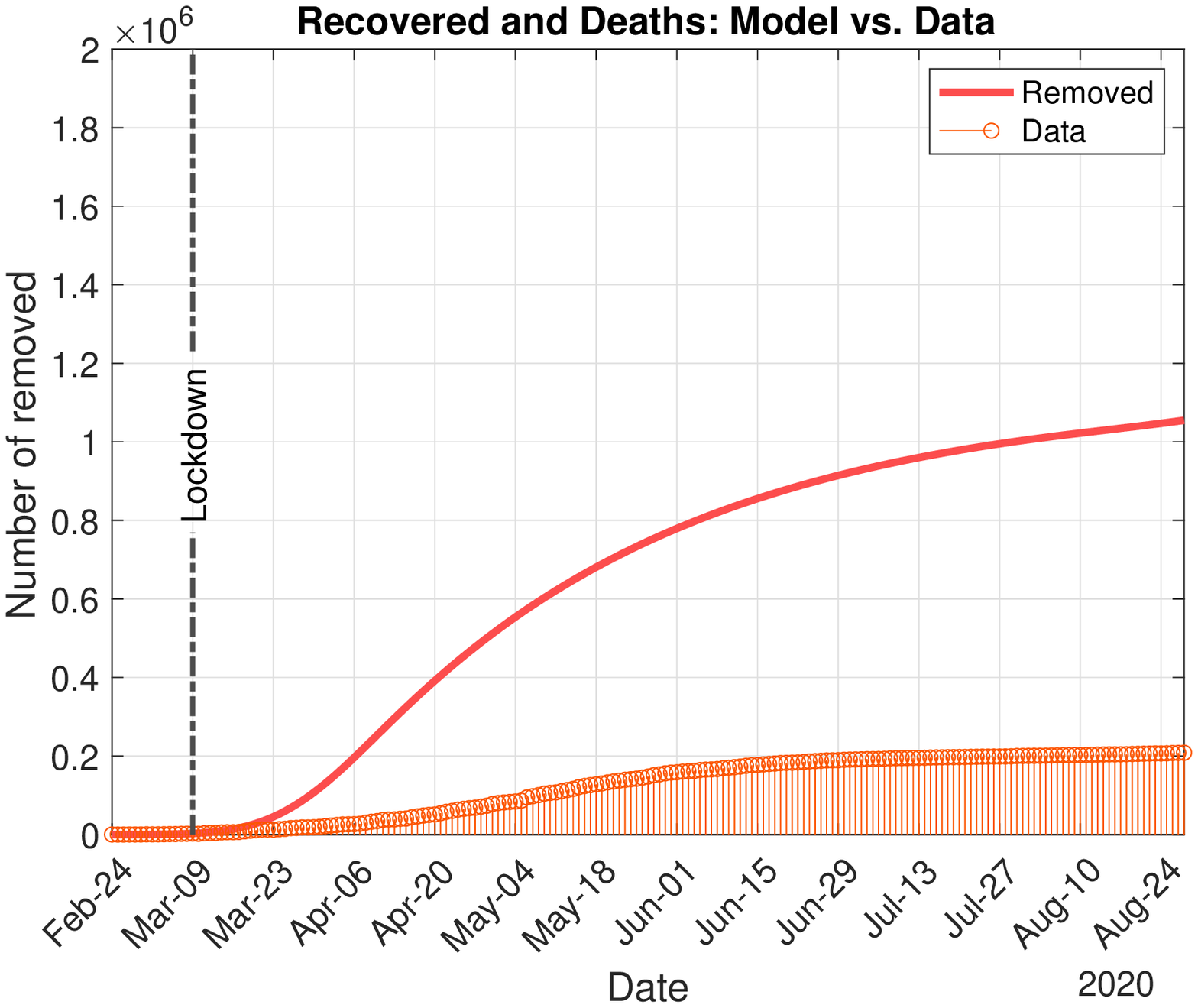}
            \caption{{\small Recovered and deaths: model vs. data\cite{pcm_dpc}.}}    
            \label{fig:italyR_data2}
        \end{subfigure}
        \caption{Model vs data: active cases (top), i.e. symptomatic and asymptomatic against hospitalised and isolated, respectively, and removed against the total cumulative recovered and deaths (bottom) on account of the Istant seroprevalence study.}
        \label{fig:italy_data_sero}
\end{figure}

\begin{figure}[t]
    \centering
    \includegraphics[width=\textwidth]{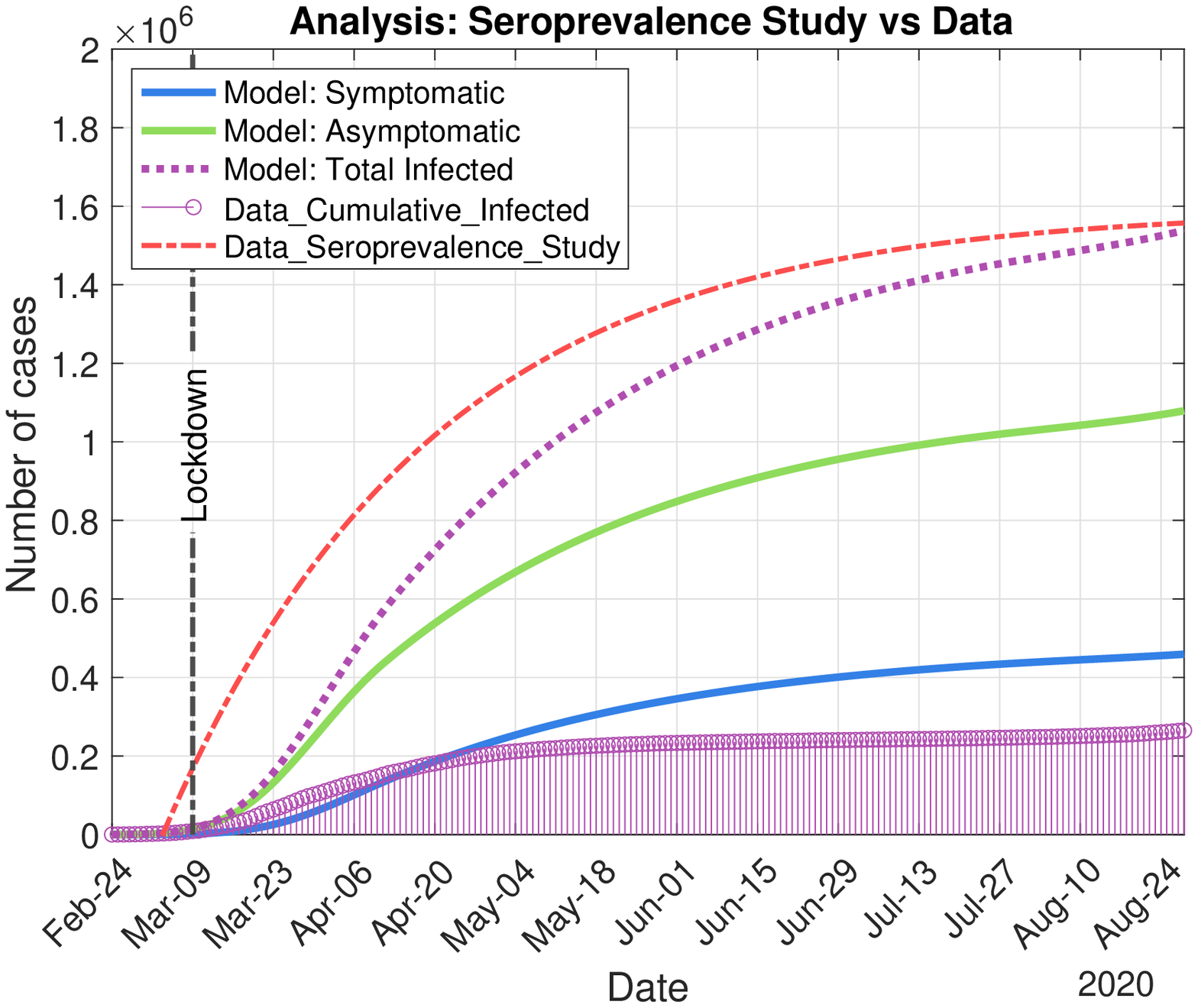}
    \caption{Analysis: seroprevalence study vs data\cite{Istat_2020}.}
    \label{fig:italySeroprevalence_sero}
\end{figure}

In the second investigation, we use the seroprevalence study to fit our model and estimate its parameters. We set the initial conditions as in the previous investigation. The other parameters are set as in the following:
\begin{eqnarray*}
\gamma&=&0.46952, \; \sigma=0.065501, \; \bar k_1 = 0.99209, \\
\lambda&=&0.48521, \; \mu=0.15004, \;  \bar k_2 = 0.65056,\\
\alpha&=&0.050017.
\end{eqnarray*}

During the days  following the local and national quarantine, we lower the values of $\bar k_1$ and $\bar k_2$ to $0.1916$ and $0.0478$, respectively, and then we account for the increased connectivity during August by setting them to $\bar k_1 = 0.2738$ and $\bar k_2 = 0.0971$. The basic reproduction number is calculated as $\mathcal R_0 = 4.9432$ at the beginning of the pandemic, and $\mathcal R_0 = 1.2490$ at the end of August. As it can be seen in Fig.~\ref{fig:italy_data_sero}, our model matches hospitalised infected very accurately, but it suggests a higher number of asymptomatic and removed to balance for matching the value of the seroprevalence study. As it is done in the previous case, we interpolate the value of the seroprevalence study by using an exponential regression as depicted in red in Fig.~\ref{fig:italySeroprevalence_sero}. We chose the parameters such that our estimation of the cumulative infected, i.e. the purple dotted curve, matches the predicted infected from the seroprevalence study. In accordance with Theorem~\ref{th3}, the ratio is in this case $\bar{\tilde I} = 0.2446$, which corresponds to the ratio between $\bar I$ and $\bar A$. We explicitly calculate the total number of people that have contracted the disease through our model by subtracting the confirmed deaths from the Removed state. We estimate a total of $8.48*10^5$ individuals who contracted that disease and are currently healthy. Most of these individuals were undetected because of being asymptomatic.

\subsection*{Complex Networks: Model vs Data}
Now, we use the proposed structured model, namely system~(\ref{eq:modCN}), to investigate the outcome of an event such as the school opening in September. We set the initial conditions as in the data\cite{pcm_dpc}, where we take the regional data and set different parameters of connectivity $\psi_{1,k}$ and $\psi_{2,k}$ depending on the region and the corresponding exposure to the virus in Italy. We set the general parameters of the model as
\begin{eqnarray*}
\gamma&=&0.49952, \; \sigma=0.05050, \alpha=0.03351, \\
\lambda&=&0.59952, \; \mu=0.15044.\\
\end{eqnarray*}

The population in each class is split according to the discretised version of the Watts-Strogatz used in the numerical analysis. The discretised distribution is the same as the actual population of each region in Italy: $[0.022\, 0.009\, 0.032\, 0.096\, 0.073\, 0.02\, 0.098\, 0.026\, 0.166\, 0.025\, 0.005\, 0.009\, 0.009\, 0.072\,$ $0.068\, 0.027\, 0.083\, 0.062\, 0.015\, 0.002\, 0.081]$, where each value is the portion of the whole population in Italy to the corresponding region ordered alphabetically, i.e. Abbruzzo for number 1, Basilicata for number 2 and so on. As in the previous case study, we gradually lower the values of $\psi_{1,k}$ and $\psi_{2,k}$ around the lockdown date and the following few days. Then, we fit our model with the data until August 29th, and then simulate what would happen in the event that schools open in mid-September. We start raising the connectivity values in correspondence of mid-late August, to account for the tourists coming to Italy and for the citizens returning from abroad. We increase these values further in correspondence to the opening of schools in mid-September. It is worth noting that the increase is proportional to the value of $\psi_{1,k}$ and $\psi_{2,k}$, namely we increase these parameters of $1\%$ of their actual value at time $t$, for each class of connectivity $k$. Therefore, regions with a higher connectivity (taken from fitting the model to the data) would have a higher increase.

\begin{figure*}[h!]
    \centering
    \includegraphics[width=\textwidth,height=1.2\textwidth]{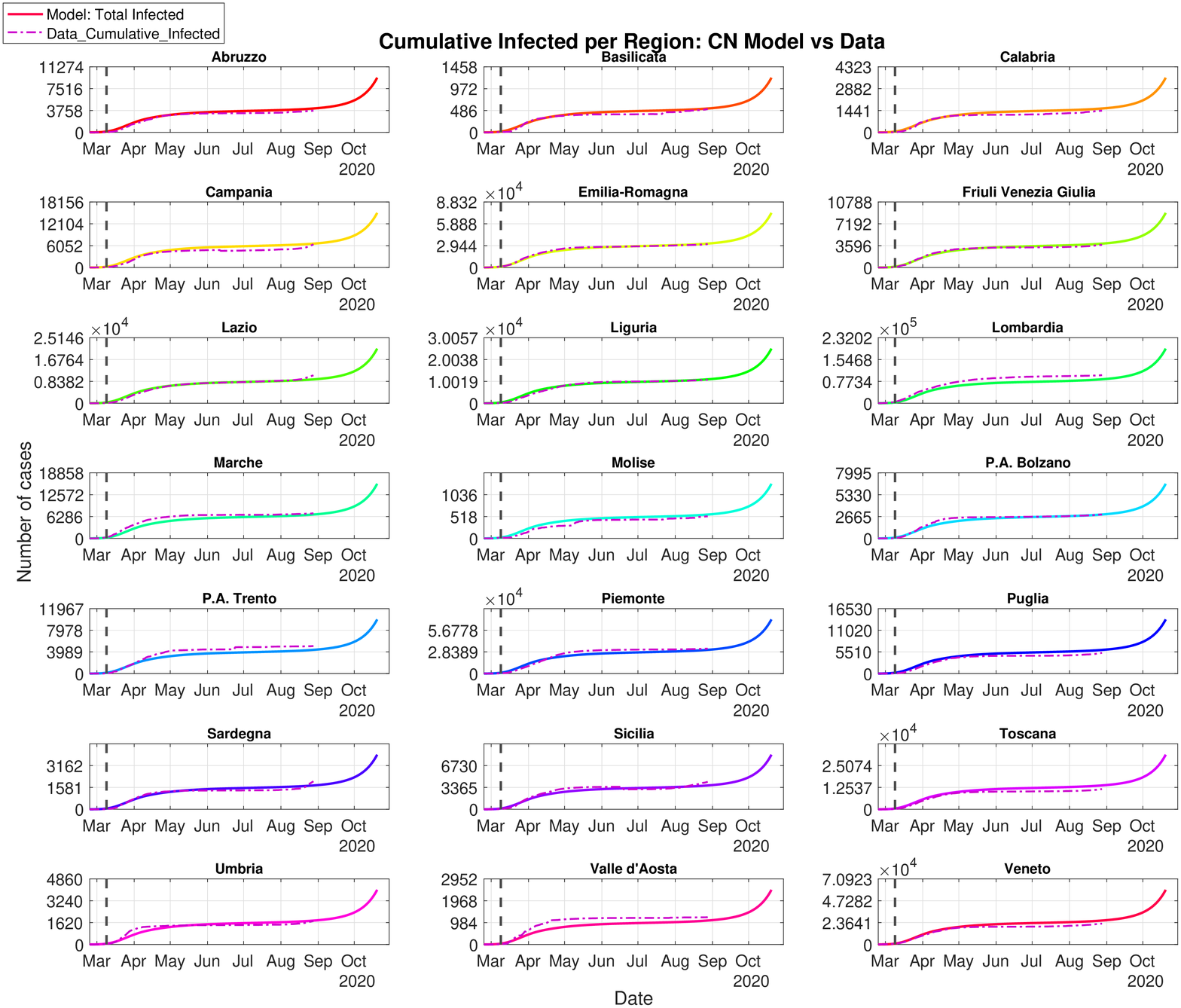}
    \caption{Total cumulative infected: model vs data. The data are regional\cite{Istat_2020}. The parameters of connectivity are set to slowly increase in the time window that correspond to the holiday (mid-August) and an additional increase in conjunction with the opening of schools (mid-September).}
    \label{fig:italyCI_region}
\end{figure*}

As it can be seen in Fig.~\ref{fig:italyCI_region}, our model captures the evolution of the cumulative infected for all regions with an error of 1\%-3\%. It is worth noting that this multi-population scenario is very difficult to fit with the data as we consider a general interaction model and not a selective interaction model, in the sense that individuals in one region interact with individuals in other regions by means of $\theta_1$ and $\theta_2$. The increase in connectivity due to holidays and return to school can take big proportions of infections, as it is already the case in France during late August. We therefore suggest caution to keep the situation under control.

\begin{figure}[h!]
    \centering
    \includegraphics[width=\textwidth]{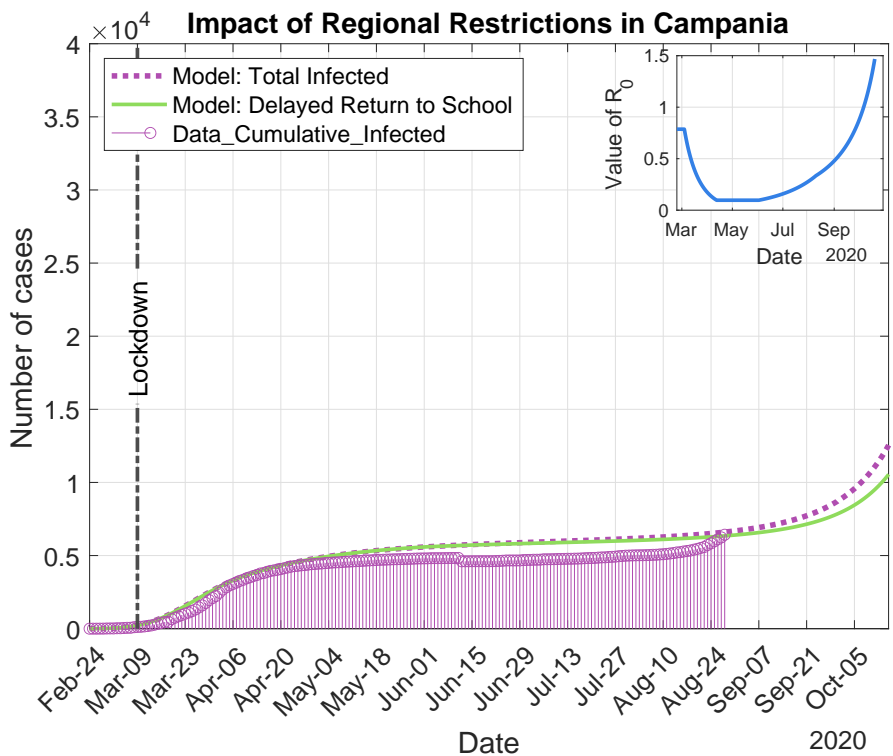}
    \caption{Scenario of delaying the return to school by one week for Campania and evolution of $\mathcal R_0$ for this region.}
    \label{fig:italy_campania}
\end{figure}

Finally, on account of the decision of the president of region Campania to delay the opening of schools in the region, we investigate the impact of such a scenario. Figure~\ref{fig:italy_campania} depicts the evolution of system~(\ref{eq:modCN}) as in Fig.~\ref{fig:italyCI_region} only for the region Campania (purple dotted line), but we have also included the evolution of cumulative infected in the scenario where schools are opened with a delay of 1 to 2 weeks by limiting the increase of the parameters $\psi_{1,k}$ and $\psi_{2,k}$ over mid-September. It is interesting to note that delaying the start of schools in just one region has a marginal benefit without closing the inter-regional movements. Indeed, while this lowers the probability of transmission in Campania from students, people travelling to this region for work from other regions that do not have the same restrictions have little to no impact without a national plan. We have also plotted the time evolution of $\mathcal R_0$ for this region  in the top-right corner: as it can be seen, the value of the basic reproduction number goes above 1 in September without further restrictions.

{\section*{\LARGE Discussion}}\label{sec:conc}
{Our study highlights the relevance of heterogeneous interactions in spreading SARS-CoV-2 while emphasising the threat of asymptomatic individuals {yet not detected and therefore not being isolated. In the asymptomatic category, our} work includes people without any symptoms or for paucisymptomatic. The Istat seroprevalence study for the Italian case guided our data-driven modelling as well as official data from Protezione Civile. We can see some differences between the model and data when keeping the parameters of infection constant and updating the connectivity measure only. This provides useful insight on the need to consider where and when to test due to contact with asymptomatic infected.}

{In Theorem~\ref{th1}, we carry out the stability analysis of the system, and link the value of the basic reproduction number $\mathcal  R_0$ to the stability of system in feedback form. This result allows us to discuss what the requirements are for the pandemic to end, mostly in relation to effective containment policies modelled via the measures of interaction. Indeed, an increase of $\bar k_1$ and $\bar k_2$ would make the value of $\mathcal R_0$ greater than 1, turning the system unstable for a potential second wave. Our analysis is refined with the results in Theorem~\ref{th2}, where we provide the limit values obtained from asymptotic calculations where the connectivity does not change after $t > t_0$, for a given $t_0 > 0$. In the next results, in Theorem~\ref{th3} we provide an expression for a nonzero epidemic prevalence, which is another way of defining the total number of infected at the end of the epidemic. In Theorem~\ref{th4}, we focus on the ratio between asymptomatic and symptomatic infected and show that this ratio is asymptotically stable.}

{To the best of our knowledge, our structured model is the different from others because we model the heterogeneous interactions in the population by means of a small world complex network in such a way that we can determine the interactions of asymptomatic and symptomatic individuals separately. A physical interpretation of Theorem~\ref{th6} in the context of the COVID-19 pandemic follows: through a variety of restrictions, local and national governments limited the number of contacts between individuals in the population and this can limit the spread of the virus, with the aim of slowing down the number of people that need intensive care. The issue with releasing the lockdown can be found in more interactions especially with people that do not show symptoms (modelled through $\psi_{1,k}$). The above result gives a value of $\mathcal R_0$ for each class of connectivity for the whole system, which translates to different restrictions in different parts of a given country to limit the localised outbreaks. The final result, namely Theorem~\ref{th7}, extends the value of the epidemic threshold in the heterogeneous case with mean $\langle f \rangle$ and parameters of connectivity $\psi_{1,k}$ and $\psi_{2,k}$ for each class of connectivity~$k$.}

{Our case study provides two clear messages. When we use the Istat seroprevalence study and fit our model with the official data, we can see a plausible evolution of the number of cumulative infected in the early stages of the pandemic. The number of asymptomatic is clearly underestimated in the official data and their role is crucial in that they can undermine the stability of the system and force another wave. When we look at the regional level, we agree with other studies that investigated similar scenarios: due to the nature of the general interaction model that we propose, uncoordinated control measures between different regions do not provide substantial benefits to contain the disease, unless the movements across regions in Italy (and analogously in Europe) are suspended, which is detrimental for workers and the economy in general.}


{\subsection*{Contributors}
LS, DB and PC contributed to the formulation of the model, the analysis and results. LS was responsible for the data acquisition, model fitting and figures. APM was responsible for the initial estimate of the parameters. All authors were responsible of the literature search. The first draft of this article was written by LS. All authors critically reviewed the first draft, and approved the final version and agreed to be accountable for the work.}

{\subsection*{Declaration of interests}
We declare no competing interests.}

{

\end{document}